\documentclass[a4paper,12pt]{article}
\usepackage[utf8]{inputenc} 
\usepackage[T2A]{fontenc}      
\usepackage[russian]{babel}    

\usepackage{amsmath,amssymb}
\usepackage[dvips]{graphicx}

\renewcommand{\baselinestretch}{1.5}
\begin{document}

МОДЕЛИРОВАНИЕ  РОСТА МОНОКРИСТАЛЛОВ: 

ТЕПЛОМАССООБМЕН 

А.И. Жмакин

ФТИ им. А.Ф.Иоффе РАН

СПб филиал МСЦ РАН

 Представлен обзор современного состояния моделирования роста объемных монокристаллов полупроводников и диэлектриков из расплава и газовой фазы в рамках моделей механики сплощной среды. Рассмотрены механизмы теплообмена теплопроводностью, конвекцией, излучением, показана необходимость корректного описания оптических свойств кристаллов (полупрозрачность, наличие диффузно и зеркально отражающих поверхностей), обсуждены вопросы характеризации качества монокристаллов, проблемы верификации моделей, особенности программного обеспечения, предназначенного для фундаментальных и прикладных исследований.

Ключевые слова: моделирование, монокристалл, рост из расплава,  из газовой фазы, теплообмен, турбулентность, верификация


\tableofcontents
\section{Введение}
\label{ch_m1:intro}

Впервые синтетические рубины получил в 1877 Э. Фреми, автор первой монографии о росте кристаллов 
 ({\it Synthesis of rubies}, 1891) \cite{sapph}. Его ученик О. Вернейль  в 1902 г. начал, используя водородно-кислородную горелку, производство
рубина и сапфира (сейчас так выращивают, например, 
кристаллов  SrTiO$_3$ \cite{srti}).
Столетие спустя  производство кристаллов оценивалось в 20 000 тонн в год \cite{2s}. 

Кристаллы выращивают из расплава, раствора,  паровой  фазы, изредка монокристаллы получают преобразованием  твердой фазы \cite{ss1,ss2} - переплавлением поликристалла путем поглощения одним зерном соседних или ростом нового зародыша за счет поликристаллической матрицы \cite{sapph}.
Рост из расплава обеспечивает наибольшую скорость --- до десятков сантиметров в час \cite{vegad}. Так получают полупроводники, оксиды, флуориды, теллуриды. Важнейшими являются кремний \cite{shimura} и сапфир \cite{sapph,bor}. Высокотемпературные сверхпроводники выращивают из нестехиометрического расплава 
 \cite{ss2,ss4}.

Вытягивание из расплава (метод Чохральского\footnote{
Сам польский ученый Ян Чохральский рассматривал эту технику не как метод выращивания, но как ``новый способ измерения скорости кристаллизации''; 
рост полупроводниковых кристаллов германия и, позднее, кремния вытягиванием из расплава был разработан Г. К. Тилом и Дж. Б. Литтлом в 1950 г. \cite{2,zuleh}. В 2012 г. Сейм Польши провозгласил 2013 Годом Чохральского.
}) --- основной промышленный способ получения монокристаллов кремния диаметром до 450 мм  \cite{vegad}
(и еще больших --- до  700 мм в диаметре --- сцинтилляционных кристаллов галидов  \cite{stri2})\footnote{Более 95\% высококачественных ({\em semicondictor grade}) монокристаллов кремния \cite{nacke} и около  45\%  кристаллов более низкого качества для солнечных батарей  ({\em solar grade}) \cite{solar} производятся методом Чохральского.
 Заметим, что кристаллы кремния, полученные методом зонной плавки, в котором высокая чистота кристалла достигается отсутствием контакта расплава с тиглем,
обеспечивают большее время жизни неосновных носителей заряда и тем самым более высокую эффективность солнечных элементов \cite{lan}.}.
Более полувека назад освоен рост бездислокационных монокристаллов кремния \cite{dash}.
При выращивании полупроводниковых соединений $A_3B_5$ для  предотвращения испарения летучего компонента из расплава  вводится дополнительный жидкий слой  (т. н. "`инкапсулированный"' рост) \cite{rud04}.
Среди других методов следует упомянуть метод Бриджмена, разработанный в 1924-1936 гг. американцем П. В. Бриджменом, немцем Д. Стокбаргером и русскими учеными И. В. Обреимовым, Г. Г. Тамманом и А. В. Шубниковым, метод Киропулоса \cite{kyr2} и его модификацию, предложенную М. И. Мусатовым в ГОИ, 
метод вытягивания профилированных кристаллов А. В. Степанова. 
Условием роста из расплава является отсутствие разложения и твердофазных переходов (как, например,  $\alpha$-$\beta$ трансформация в кварце  $SiO_2$). 
 
Из расплава нельзя вырастить кристаллы широкозонных полупроводников, таких, как карбид кремния SiC  и нитриды III группы (AlN, GaN, InN и их твердые растворы). Уникальные свойства нитридов III группы
\cite{prop2,epi} 
делают их лучшим материалом для оптоэлектронных приборов в видимом и УФ диапазонах \cite{orton}, включая источники освещения 
 \cite{uv,zuk,schub}, а также мощных и СВЧ приборов, работающих при высоких температурах   и в агрессивной среде \cite{tran2,peart}. 
Чтобы использовать потенциал этих материалов, требуются нативные подложки высокого качества\footnote{Требования к подложке следующие \cite{z10aln}: согласованность по постоянной решетки в двух направлениях (рассогласование в вертикальном направлении ведет к генерации дефектов путем смещения эпитаксиальных слоев (ЭС), включая обрашение границ доменов и дефекты упаковки  \cite{epi}); согласование по коэффициенту линейного теплового расширения;  
химическая совместимость; большой размер ("размер, приемлемый для промышленности"  \cite{cry_is} --- не менее 2$^{\prime \prime}$); доступная цена. Иногда желательны дополнительные свойства, например, спайность для  лазерных диодов или высокое электрическое сопротивление для полевых транзисторов. Подложка определяет кристаллографическую ориентацию, полярность, политип и морфологию выращиваемых ЭС; различия в химическом составе подложки и ЭС могут привести к загрязнению последнего.
}: гомоэпитаксия обеспечивает меньшую, чем гетероэпитаксия, плотность дефектов \cite{cao07}.

Кристаллы AlN, обладая наибольшей среди известных пьезоэлектриков поверхностной и объемной акустической скоростью, малым температурным коэффициентом задержки и высоким пьезоэлектрическим коэффициентом \cite{bu04}, идеальны также для приборов, использующих поверхностно-акустические волны (ПАВ)\footnote{Интересны даже поликристаллические пленки 
AlN (свойства упругих волн в поликристаллах, однако, зависят от размера и ориентации зерен), поскольку соответствующие ПАВ приборы могут работать на частотах выше 1 ГГц  \cite{epel_saw} при температурах до 1150$^{\circ}$C \cite{fritze}, в отличие от традиционных пьезоэлектриков (в кварце  $\alpha SiO_2$ 
происходит фазовый переход при  573$^{\circ}$C, а ниобат лития $LiNbO_3$ разлагается при 300$^{\circ}$C). AlN рассматривается также как материал для интеграции полупроводниковых и ПАВ приборов \cite{cleland}.}.  

Из-за недостатка нативных подложек  (GaN, AlN) используются чужеродные (GaAs, AlAs, GaP, ZnO, MgO, LiGaO$_2$,  SiO$_2$,  ZrO$_2$, Mo и т. п. \cite{prop2,sub}), что  снижает качество ЭС (рост плотности проникающих дислокаций (ППД), появление дефектов упаковки,  трещины  \cite{speck2}). Лучшие - подложки из сапфира\footnote{
Сапфир Al$_2$O$_3$ привлекателен с кристаллографической точки зрения: элементарная ячейка GaN, выращенная на с-плоскости (0001), повернута на  30$^{\circ}$ вокруг оси c по отношению к ячейке сапфира, а ось ($1\bar{1}\bar{0}0$) GaN параллельна оси ($1\bar{2}\bar{1}0$) сапфира \cite{prop2}.
Однако ЭС, выращенные на сапфире, уступают по качеству выращенным на карбиде кремния (6H-SiC,  4H-SiC) в силу значительно большего рассогласования постоянных решетки и коэффициентов линейного теплового расширения.
Изолируюшая природа сапфира исключает вертикальную архитектуру чипа: биполярные приборы на подложках $Al_2O_3$ должны иметь анод и катод в одной плоскости и потому чувствительны к эффекту концентрации тока
(current crowding) в окрестности электродов. Это явление ограничивает эффективность светодиодов (СД), биполярных транзисторов и диодов Шоттки \cite{paskova10}.  Концентрация тока приводит к перегреву, формированию горячих пятен и выходу прибора из строя \cite{bul12,spr2}. Другим недостатком сапфира как подложки для нитридных СД является меньший, чем у нитридов, коэффициент преломления, ведущий к волноводному эффекту и снижению эффективности вывода света \cite{zhm11}.  
Тем не менее, около  80\% нитридных СД производится на сапфировых подложках диаметром 2'' в силу их низкой стоимости и доступности  \cite{2in}. 
} и карбида кремния.

Хотя первые кристаллы  AlN синтезированы 150 лет назад Ф. Бриглером и А. Гюнтером (см. \cite{sitar}), а игольчатые кристаллы GaN 
Р. Джуза и Х. Ханном в 1938 г. \cite{prop2}, рост кристаллов 
AlN, GaN и AlGaN остается трудным \cite{denis}; монокристаллы InN не получены до сих пор --- термическая неустойчивость нитридов III группы возрастает с номером в периодической системе Д. И. Менделеева   \cite{ipap}.

Монокристаллы нитридов III группы нельзя вырастить из расплава при атмосферном давлении из-за разложения при 1150 K \cite{am}  вследствие сильной энергии связи молекулы азота и низкой свободной энергии кристалла \cite{g1}: GaN разлагается на металлический галлий и газообразный азот \cite{cg}. 
Конгруэнтное плавление нитрида галлия недавно осуществлено при давлении 6 ГПа \cite{utsumi}.

Нитриды III группы выращивают из раствора или из газовой фазы. Растворимость азота в галлии  низка \cite{g3} (в 
пересыщенном растворе образуются пузырьки N$_2$  \cite{g1}), но 
 может быть увеличина  растворением радикалов азота ---  аммиак предпочтительней как окружающий газ  \cite{kawa}.
Кристаллы  GaN, выращенные при сверхвысоком давлении (High Nitrogen Pressure Solution Growth) \cite{por,g4}, имеют рекордно низкую ППД --- $10^2cm^{-2}$. Получены игольчатые и объемные кристаллы размером 1 см и 1мм  соответственно \cite{bock}.
Выращены также кристаллы  (Al,Ga)N с содержанием алюминия   0.22 - 0.91, размер самого крупного составлял 0.8 х 0.8 х 0.8 мм  \cite{bel1}. 

От экстремальных параметров этого метода удается уйти в аммонотермальном росте \cite{5,13} и методе щелочных металлов \cite{g19,g20}. Первый 
принадлежит широкому классу сольвотермальных методов и 
подобен росту аметистов в природе \cite{ameth} и гидротермальному методу выращивания кварца \cite{quartz}.
Низкая скорость роста компенсируется одновременным ростом множества кристаллов (более 100 в случае 
ZnO и более 2000 в случае кварца). Сотни различных кристаллов выращиваются сольвотермальным методом, но единственный пример промышленного роста широкозонных полупроводников --- рост оксида цинка \cite{cal13}  при давлении кислорода 50 атм \cite{cal9}.

В этом методе полярный растворитель (например,  аммиак в аммонотермальном) образует метастабильные промежуточные продукты с растворенным веществом (нутриентом). Растворимость нутриента повышают минерализаторы. Так, 
GaN не удается вырастить в растворе  галлия ---  необходимо добавить литий как переносчик либо кислотный или основной минерализатор \cite{calla} - NH$_4$X (где X= Cl, Br, I) \cite{5,13} или GaI$_3$ (вместе с CuI или LiI) \cite{5}. 

В методе щелочных металлов молекула азота, абсорбированная на поверхности раствора 
Ga-Na, получает электроны от натрия, что ослабляет связи в молекуле и вызывает диссоциацию N$_2$ на два радикала при более низких температуре и давлении \cite{g19}. 
Замена натрия на литий позволяет снизить давление до 1-2 атм \cite{g20}. 

Скорость роста нитрида галлия из раствора не превышает 2$\mu$/час \cite{fukuda}).
Не следует ожидать развитие этой технологии до промышленного уровня в обозримом будущем.
 
Сублимационный\footnote{Сублимация --- образование паровой фазы из твердой. Сублимационный рост (неявно предполагается, что обе фазы --- одно и то же вещество) обеспечивается бездиффузионным переносом пара: сублимация источника и конденсация на затравке происходят конгруэнтно --- нет изменений состава, нет посторонних компонентов - перенос обеспечивается т.н. ветром Стефана (дрейфовый перенос \cite{s6}) и скорость роста максимальна \cite{brink}.  Строго говоря,  AlN и SiC не \textit{сублимируют}, а \textit{разлагаются}.
} рост (также - конденсация из паровой фазы, модифицированный метод Леви), разработанный Ю. М. Таировым и В. Ф. Цветковым \cite{tai}, используется для получения кристаллов SiC и AlN \cite{sitar,sitar10,z11,z12}, ZnO \cite{ro06}, волокон  AlN  \cite{fib}, других нитридов, например, нитрида титана \cite{titan}.  Сублимационный рост кристаллов GaN менее успешен \cite{kallinger} --- равновесное давление азота над поверхностью GaN на шесть порядков выше, чем над  AlN \cite{freitas10}.

Требования к монокристаллам  ужесточаются: так, кристаллы сапфира диаметром 14 см  рутинно выращиваются методом температурного градиента (TGT) \cite{tgt}.
Нужны, однако, оптически однородные кристаллы сапфира диаметром 35 см (для лазерного интерферометра в обсерватории гравитационных волн \cite{ligo,ligo1}).
 
Проектирование т. н."'горячей зоны"'(нагревательных и теплоизолирующих элементов  \cite{lan})  трудно. Простое масштабирование  \cite{2,bagd} не годится из-за нелинейности процессов переноса: так, число Рейнольдса $\rm Re$ линейно зависит от характерного размера, число Грасгофа $\rm Gr$ --- как третья степень, а число Марангони $\rm Mn$  не зависит; конвекция расплава становится неустойчивой при росте размера кристалла \cite{zeng}.  Моделирование может помочь, если  предсказания достоверны.

Модели используют и для управления процессом роста в реальном времени \cite{derby1} --- т.н. предсказательные системы контроля \cite{mpc2} или основанные на моделях системы обратной связи \cite{rud07}. В  методе Чохральского, например, необходимо выдерживать постоянным диаметр кристалла --- для коррекции можно использовать измерения высоты кристалла,  формы мениска и т. п. \cite{jang}.

В статье дан обзор макроскопического численного моделирования тепломассообмена при выращивании монокристаллов. Эффекты кристаллизации на  микро- и  наномасштабах \cite{chern1,chern2,weeks,raffi} не рассматриваются, так же как и моделирование роста дендритов и эволюции квазиравновесной двухфазной зоны  
\cite{10,11,dend}.
\section {Математические модели}
Основным процессом при росте кристаллов является теплообмен, кристаллизация чистого вещества определяется только полем температуры. Есть исключения: эпитаксия в реакторах с горячей стенкой 
 \cite{3}, рост из раствора \cite{vartak}, включая кристаллизацию белков\footnote{
Кристаллы нужны для исследования структуры методами рентгеновской кристаллографии, рассеяния нейтронов, электронной микроскопии. Банк данных в 2014 г. содержал $10^5$ 3D структур белков и нуклеиновых кислот.
К сожалению, нет доказательств идентичности структуры белка в кристалле и в "`рабочей конформации"' --- в растворе.
Значение воды ("21-й аминокислоты"  \cite{levy})  для функционирования белка нельзя переоценить \cite{rand}.  
Иногда нормальное функционирование белковой молекулы требует покрытия ее поверхности бесконечным  (в терминах теории протекания) кластером воды
 \cite{smolin}. 
}, жидкофазная эпитаксия  InGaAsP на движущуюся подложку  \cite{4}. 

Большинство неорганических кристаллов выращивают при высоких температурах --- радиационный теплообмен является основным.
Теплопроводность и конвекция определяют локальное распределение температуры и тем самым скорость роста и морфологическую устойчивость. Рост кристаллов затрагивает широкий спектр масштабов: от нанометров через мезомасштабные структуры (дефекты, вторичные включения  и т. п.) к макроскопическим монокристаллам. Представление о диапазоне размеров установок дает сравнение двух родственных процессов --- аммонотермального роста GaN в реакционной трубке и роста кварца в автоклаве с производительностью до 3 тонн (Табл. 1).
Значительно варьируются и свойства жидкости. Число Прандтля меняется от 0.01 для расплавов Si и Si-Ge \cite{abba} до десятков для расплавов оксидов, что определяет качественные отличия в конвекции \cite{crn08}, в частности, неустойчивости течения в методе Чохральского приводят к спиралевидным формам монокристаллов некоторых оксидов  
(DyScO$_3$, SmScO$_3$).

Моделирование роста кристаллов требует решения сопряженной задачи 
 \cite{lan,1,96,zhm04}. Так, модель сублимационного роста SiC включает резистивный или СВЧ нагрев, теплообмен теплопроводностью, конвекцией, излучением, сопряженный массообмен в газовом зазоре и в пористом источнике, гомогенные химические реакции и гетерогенные реакции на стенках установки и поверхности гранул источника, 
формирование упругой деформации и дислокаций в растущем кристалле, образование поликристаллических  депозитов на стенках ростовой ячейки, эволюцию формы монокристалла и депозитов с учетом огранки кристалла, образование графитовых или кремниевых включений в кристалл, эволюцию состава и пористости источника \cite{bogd,14,vr1}.

О роли различных механизмов  судят по безразмерным критериям подобия (Табл. 2). 
Делать это следует с осторожностью: например, часто преобладание естественной или вынужденной конвекции оценивается по числу Ричардсона 
$\rm Ri = Gr/Re^2$ \cite{quasi1}; однако смешанное течение может быть неединственным, а реализация конкретной структуры определяться историей процесса  \cite{41}.

Ключевым является расчет течения жидкости, определяемого  теплообменом в установке 
 ({\em furnace-scale}, {\em process-level} \cite{derby1}). 
Поскольку характерное время изменения формы кристалла (около минуты  в методе Чохральского) велико в сравнении с  гидродинамическим, часто используют квазистационарный подход --- решается ряд стационарных задач, соответствующих разным стадиям роста.
Нестационарное моделирование необходимо при изучении эволюции примесей (C, O) и  точечных дефектов (межузлия, вакансии) в кремнии 
\cite{kulk,tal,ammon}, а также   эволюции  дислокаций, когда требуется тепловая история процесса (включая охлаждение) \cite{transient,tr1}.

Цель численного моделирования --- объяснение и предсказание. Оно не заменяет, но дополняет эксперимент, предоставляя информацию, которая экспериментально может быть получена только косвенно. С другой стороны, модели опираются на опытные данные (материальные свойства и т. п.) и требуют верификации. В англоязычной литературе используются, часто как синонимы, термины  {\em modelling}  и {\em simulation}, 
хотя первый рекомендуется относить к разработке модели, а второй --- к ее использованию  \cite{7,raabe}. 

Модель связывает спецификацию процесса с результатом 
(Табл. 3) --- это {\it прямая} задача. 
На практике полезнее иная формулировка: как следует изменить конструкцию или параметры процесса, чтобы повысить качество кристалла или снизить стоимость  \cite{chen08}. Интуитивный метод проб и ошибок \cite{trial} и еще более простой "`слепой поиск"' \cite{8} требуют значительных средств и времени. Можно поставить {\em обратную задачу}, указав, какие параметры могут меняться и каков критерий успеха ({\em inverse modeling} \cite{kurz},  {\em inverse engineering} \cite{raabe}). Обратные задачи некорректно поставлены  - необходима дополнительная информация для регуляризации задачи  \cite{9} (ограничение класса решений, например, требование гладкости \cite{zhm90}).

Макроскопическое моделирование роста кристаллов основано на моделях сплошной среды. В случае  объемных кристаллов нельзя, в отличие от ЭС \cite{ipa,zhm90_m}, считать расчетную область фиксированной: необходимо либо использовать подвижные расчетные сетки, либо периодически проводить регенерацию. Последний подход особенно привлекателен для квазистационарного моделирования.

\paragraph{Глобальная и локальная задачи}
Часто решение глобальной задачи служит источником  граничных условий для меньшей области, в которой используются более сложные физические модели (например, полупрозрачность кристалла) или модели более высокой размерности (трехмерное  течение в осесимметричной ростовой установке \cite{37,z8si,z10,mart14}).  
В последнем случае нельзя ограничить 3D область расплавом (даже расширяя ее за счет части кристалла \cite{derby1}): осесимметричные граничные условия искусственно ограничивают турбулентное течение. 
Например, при моделировании роста 
 Si методом Чохральского 3D область должна включать кварцевый и графитовый тигли, а в ряде случаев --- и часть газовой области  \cite{37,38,kak13} --- 
распределение температуры на границе расплав/тигель существенно нестационарно и далеко от осесимметричного. Неоднородность теплового потока на этой границе - источник колебаний концентрации кислорода в растущем кристалле.
Т.о., проблема подразделяется на две
\begin{description}
\item[Глобальный ({\em furnace-scale}) теплообмен]{Теплообмен во всей установке в области  $\mathcal{D}_{global}^{2D}$ }
\item[Процессы в горячей зоне]{Конвекция и тепломассообмен в расплаве и ближайших элементах ростовой установки $\mathcal{D}_{local}^{3D}$}
\end{description}
Самосогласованное решение достигается итерациями решений в областях  
 $\mathcal{D}_{global}^{2D} \setminus \mathcal{D}_{local}^{3D}$ и $\mathcal{D}_{local}^{3D}$, связанных тепловыми условиями на границе  $\partial \mathcal{D}$. Однонаправленное сопряжение (например, задание температуры ампулы, содержащей расплав и кристал в реальном процессе, из расчета для {\em пустой ампулы} \cite{mar14}) обычно неприемлемо.
\subsection{Теплообмен}
\subsubsection{Теплопроводность}
Стационарное распределение температуры удовлетворяет уравнению  
\begin{equation} 
\nabla \cdot \left( -\lambda_{solid} \nabla T \right)=Q, 
\end{equation}
где  $Q$ --- источник тепла, например, СВЧ нагрев.
Коэффициент теплопроводности $\lambda_{solid}$ анизотропен для кристаллов (степень анизотропии невелика, а необходимость 3D
рассмотрения определяется ориентацией кристалла и геометрией установки)  и блоков из пиролитического графита, имеющего слоистую структуру с упорядоченным гексагональным расположением атомов углерода в плоскости; отношение коэффициентов теплопроводности  составляет от 100 до 400.
 
Элементы установки могут являться пористой средой.
Пористый источник используется при выращивании кристаллов GaN 
из раствора  Li-Ga-N \cite{12}, в аммонотермальном методе \cite{13a,13}, 
сублимационном росте SiC и AlN \cite{14,15}. Гранулированные или волокнистые материалы служат теплоизоляцией. Эффективный коэффициент теплопроводности $\lambda_{eff}$, зависящий от давления и температуры, суммирует вклады излучения, теплопроводности твердой матрицы и газовых пустот. Опытные данные скудны, поэтому необходима модель для экстраполяции $\lambda_{eff}$  за пределы исследованных интервалов давления и температуры \cite{16}.

\subsubsection{Конвективный теплообмен}
\paragraph{Течение расплава}
Разнообразие картин течения отражает многообразие движущих сил и различия в масштабах и  свойствах  \cite{19}. 
Простейшее течение в методе Бриджмена  \cite{dupret94} --- это тепловая и/или концентрационная конвекция, характеризуемые ($\beta_c$  --- концентрационный коэффициент расширения) 
тепловым
$
Gr = {\beta g L^3 \Delta T}/{\nu^2}
$
и концентрационным  
$
Gr_c = {\beta_c g L^3 \Delta c}/{\nu^2}
$
числами Грасгофа. 

Течение в вертикальном методе Бриджмена с центрифугированием  \cite{21} определяется взаимодействием сил плавучести и Кориолиса.
Во вращательном варианте \cite{22} добавляется вынужденная конвекция.  
Вынужденное течение в методе Чохральского вызывается вращением  кристалла и тигля. Сдвиговое напряжение на свободной границе расплава из-за движения газа может превышать напряжение, связанное с эффектом Марангони  \cite{23}). В инкапсулированном методе Чохральского появление третьей  жидкости несущественно усложняет расчет течения (течение инкапсулянта ламинарное), но затрудняет моделирование массообмена --- сопряженных  физико-химических процессов в трех средах.

В отличие от упомянутых методов в методе зонной плавки фазовая граница расплав-кристалл имеет сложную форму
\cite{24}: жидкий мостик образуется между двумя твердыми поверхностями, а течение определяется градиентами температуры и концентрации. Период колебаний в случае смешанного термоконцентрационного эффекта Марангони меньше (и, значит, требуется более высокое разрешение при расчете), чем в случаях, когда течение вызывается изменением только температуры (или концентрации)  \cite{minak}.

Течение расплава описывается следующими уравнениями
\begin{equation}
\dfrac{\partial \rho}{\partial t} + \nabla \cdot (\rho \vec{V}) = 0\\
\end{equation}
\begin{equation}
\dfrac{\partial (\rho \vec{V})}{\partial t} + (\vec{V} \cdot\nabla) \rho \vec{V} = -\nabla p +  \nabla \cdot \hat{\tau} +(\rho - \rho_0) \vec{g} + \vec{j} \times \vec{B} \\
\end{equation}
\begin{equation}
\dfrac{\partial (\rho C_p T)}{\partial t}) + \nabla \cdot (\rho C_p \vec{V} T) = \nabla \cdot (\lambda_{eff} \nabla T) -\nabla \cdot \vec{q}^{rad} \\
\end{equation}
\begin{equation}
\dfrac{\partial (\rho \phi_i)}{\partial t} + \nabla \cdot (\rho \vec{V} \phi_i) = \nabla \cdot (D_{\phi_i,eff} \nabla \rho \phi_i)) \\
\end{equation}
\noindent где $\rho$ --- плотность,  $\vec{V}$ --- скорость,  $p$ --- давление, $\hat{\tau}$ --- тензор напряжений, $\mu_{eff} = \mu_{mol} +
\mu_t$ --- эффективная динамическая вязкость, $C_p$ --- удельная теплоемкость, $T$ --- температура, $\phi_i$ --- доля i-й  примеси, $\lambda_{eff} = \lambda_{mol} + \mu_t/Pr_t$ --- эффективная теплопроводность,
$\vec{q}^{rad}$ --- поток тепла излучением, $D_{\phi_i,eff}$ --- эффективный коэффициент диффузии, $\vec{j}$ - электрический ток, $\vec{B}$ --- магнитная индукция.

Обычные условия прилипания для скорости,  непрерывности температуры и теплового потока ставятся на границе расплав/тигель. На свободной поверхности расплава  нормальная скорость равна 0, а тангенциальные скорости расплава и газа удовлетворяют условиям  
\begin{align}
\left(\mu_{eff} \dfrac{\partial V_{\tau}}{\partial n} \right)_1 =
\left(\mu_{eff} \dfrac{\partial V_{\tau}}{\partial n} \right)_2 +
\dfrac{\partial \sigma}{\partial T} \nabla_{\tau} T \\
(V_{\tau})_1 = (V_{\tau})_2
\end{align}
При росте  соединений $A_3B_5, A_2B_6$ и  таких полупроводников, как кремний-германий\footnote{Этот сплав позволяет подстраивать ширину запрещенной зоны и полностью совместим с  технологией кремниевых приборов \cite{sige1}. Монокристаллы SiGe выращивают также из раствора  Si-Ge методом жидкофазной диффузии \cite{lpd,lpd1}, основанном на принципе пересыщения - в отличие от роста из расплава, обеспечивающимся охлаждением. 
} \cite{bolkh,sige} необходимо предотвратить изменение состава кристалла, вызванного отторжением компонента от фазовой границы, подпиткой расплава одним из компонентов  \cite{armour}.

Естественная конвекция в расплаве в условиях микрогравитации, приводящая к неоднородному распределению примеси, связана с эффектом Марангони и/или нестационарными возмущениями (g-jitter), вызываемыми движениями экипажа и механическими колебаниями спутника. 
Амплитуда этих ускорений может на порядок превышать ускорение силы тяжести на орбите; иногда следует учитывать силу Кориолиса из-за вращения спутника вокруг оси и вокруг Земли (см. \cite{25,25a} 
и приведенные ссылки). 
Зависимость максимальной среднеквадратичной скорости от числа Экмана
Ek = $\nu/ (2 \Omega L_0^2)$ или от частоты микроускорений имеет резонансный характер.  
Вибрации можно также использовать для управления ростом как в земных условиях, так и на орбитальной станции  \cite{27}.

О характере течения ---  ламинарное или турбулентное --- судят по числу Рэлея в случае естественной конвекции и числу Рейнольдса в случае вынужденной \cite{pol}.
Так, течение турбулентно в гидротермальном росте кварца: число Рэлея, определенное по диаметру автоклава, достигает $Ra = 4 \cdot 10^{11}$ \cite{6}.
Турбулентным является течение расплава в росте кремния методом Чохральского в силу больших масштабов и малой вязкости расплава  \cite{z8si,28}.

Моделирование роста методом Чохральского имеет богатую историю
\cite{dupret94,Langlois_77,Samarskii_86,Brown_88,Muller_88}.
Первые расчеты выполнены для осесимметричного ламинарного течения.
Трехмерные расчеты, начатые в  \cite{BottaroZebib} 
и продолженные другими исследователями \cite{Seidl,Kakimoto}, показали,  
что неосесимметричное нестационарное течение формируется уже с чисел  Рейнольдса порядка  $10^4$.
Возникающие волновые структуры связывают с бароклинной неустойчивостью
(\cite{Seidl}. Эксперименты (см. ссылки в  \cite{Hurle}) также показывают, что течение расплава нестационарное и, возможно, турбулентное, для кристаллов с диаметром более 40 мм.
Авторы  \cite{Ristorcelli_a} рассмотрели свидетельства турбулентного характера течения расплава  в промышленных установках для выращивания кремния и предложили использование усредненных по Рейнольдсу уравнений Навье-Стокса.
Стандартная высокорейнольдсова $k - \epsilon$
модель турбулентности с использованием пристеночной функции для реализации условия прилипания  \cite{LaunderSpalding} впервые использована в 
\cite{Kobayashi90}.

Применение этой модели к течению расплава кремния неоправдано. Традиционные пристеночные функции получены для сдвиговых слоев с локально равновесной турбулентностью --- генерация турбулентности примерно уравновешивается диссипацией и конвективный перенос играет вторичную роль. Пристеночные слои у кристалла и тигля ламинаризованы, в них генерации турбулентности не происходит  - турбулентные пульсации переносятся  из ядра потока. 
Вращение тигля подавляет турбулентность и турбулентная вязкость становится сопоставима с молекулярной.

Кинни и Браун  \cite{KinneyBrown} использовали низкорейнольдсову  $k - \epsilon$ модель,
требующую более высокого пространственного разрешения, 
чем стандартная $k - \epsilon$ модель.
Обе модели основаны на изотропной турбулентной вязкости. Ристорелли и Лумли \cite{Ristorcelli_b} отметили, что анизотропия турбулентности важна для течения расплава кремния и указали  на  модели второго порядка. Использование уравнений для рейнольдсовых напряжений позволяет непосредственно учесть влияние плавучести и  вращения на турбулентное смешение, однако  увеличивает число дополнительных уравнений в частных производных. Ни одна из моделей второго порядка не была верифицирована для течений, где существенны все факторы, играющие важную роль в течении расплава кремния.

Трудность построения модели турбулентности - в нелинейном взаимодействии явлений всех масштабов и отсутствии спектрального разрыва, который позволил бы отделить  далекие от равновесия крупномасштабные движения от находящихся в равновесии (термализованных) мелкомасштабных  \cite{farge}.
Использование усредненных по Рейнольдсу уравнений Навье-Стокса (РУНС) не позволяет описать форму межфазной границы или распределение кислорода в кристалле кремния \cite{29,30}.

Прямое численное моделирование (ПЧМ) обеспечивает полную информацию о течении, включая пространственно-временные корреляции \cite{corr}, однако его применение возможно только для обладающих большой вязкостью течений расплавов оксидов или в областях простой формы. Утверждение  "`ПЧМ - метод исследования, а не инструмент решения инженерных проблем"'\cite{moin}  остается справедливым \cite{antt}.
При росте кристаллов кремния диаметром 300 мм числа Рейнольдса, вычисленные по скоростям вращения кристалла и тигля,  --- $Re_{crys} \approx 10^5$ и $Re_{cruc} \approx 3 \cdot 10^5$, что соответствует числам Рейнольдса по тейлоровскому масштабу\footnote{
Тэйлоровский масштаб - масштаб расстояния, на котором молекулярная вязкость существенно сказывается на динамике турбулентных вихрей; его вычисление использует некоторые корреляционные функции течения \cite{taylor}.
}  $Re_{\lambda}$  400-600 (Табл. 4) \cite{32}. 
Наивысшее значение $Re_{\lambda}$ = 1200 
получено для {\em однородной} турбулентности в периодической кубической области спектральным методом с использованием расчетных сеток,
содержащих 
$4096^3$ (($69 \cdot 10^9$) ячеек \cite{34}.  
Размер сетки для этой простой задачи при указанных выше значения 
$Re_{\lambda}$ - около  $10^9 - 10^{10}$ ячеек \cite{34,34a}. 
Еще более подробные сетки нужны для расчета переноса кислорода вследствие большого значения числа Шмидта \cite{35,36} - для кислорода в расплаве кремния $Sc \approx 10$ \cite{organ}. 

Очевидно, нельзя доверять результатам т.н. "`грубого ПЧМ"' или "`квази ПЧМ"'  \cite{quasi}, которые не обеспечивают необходимого пространственного разрешения\footnote{
Колмогоровский масштаб 
$\eta = (\nu^3/\epsilon)^{1/4}$
(где $\epsilon$ --- средняя скорость диссипации турбулентной кинетической энергии на единицу массы, а  $\nu$ - кинематическая вязкость)
считается наименьшим масштабом, который должен быть разрешен. Эта величина  мала - несколько миллиметров даже для атмосферных вихрей с характерным размером порядка километров. 
Размер ячейки расчетной сетки должен быть около 
$0.26 \eta$,  $0.55 \eta$ и  $0.95 \eta$  для разностных схем 2, 4 и 6 порядков, соответственно \cite{moin}. 
Число узлов/ячеек/элементов в схемах конечных разностей/объемов/элементов 
должно быть пропорционально числу степеней свободы турбулентного течения 
$N_{FD,FV,FE} \propto  N^{DOF}$. 
Известно, что 
$N^{DOF} \propto Re$ и  $N^{DOF} \propto Re^{9/4}$ 
для двумерных и трехмерных течений, соответственно.
По-видимому, в настоящее время наилучший подход к моделированию турбулентных течений при очень больших числах Рейнольдса --- адаптивный метод вейвлетов
\cite{farge92},
поскольку требуемое число мод 
$N^{mod}_{AWM}  \propto N^{CV}$, где  $N^{CV}$  --- число когерентных вихрей в потоке \cite{schneid}.
}, 
но, как утверждают их защитники, дают картины течения, похожие на наблюдаемые. Действительно, когерентные структуры в развитых турбулентных течениях слабо зависят от числа Рейнольдса. При росте кристаллов, однако, интересна не картина течения, а его поведение вблизи границ - кристалла, тигля, свободной поверхности.
Эта ситуация напоминает попытки в 80-е годы ХХ века использовать уравнения Эйлера для расчета отрывных вязких течений: расчеты воспроизводили общую картину течения, если точка/линия отрыва определялась геометрией задачи, но не могли дать коэффициент трения или теплопередачи.

В настоящее время оптимальными подходами к моделированию турбулентного течения расплава являются метод крупных вихрей (МКВ) 
\cite{liu12} 
и гибридные методы, сочетающие лучшие стороны МКВ и РУНС 
\cite{37,38a,ev02}. 
В МКВ явно разрешаются крупномасштабные движения, а эффект мелкомасштабных движений учитывается т.н. подсеточной (Subgrid Scale -
SGS) моделью .
В гибридном  методе \cite{38,29} подсеточная модель активируется в ядре потока расплава, а $k - \epsilon$ модель обеспечивает генерацию турбулентности вблизи твердых поверхностей.

\paragraph{Течение газа}
  Течение газа в задачах роста кристаллов ламинарное, за исключением инкапсулированного роста полупроводников A$_3$B$_5$ при высоком давлении,
	и характеризуется малыми по сравнению со звуковой скоростями - число Маха $M = V_0/a \ll 1$, но значительными изменениями температуры (и, следовательно, плотности). 
	Такие течения называются  гипозвуковыми \cite{zeyt83}.

	Использование уравнений Навье-Стокса  сжимаемого газа для исследований течений, которые характеризуются двумя сильно различающимися масштабами скорости  и, следовательно, времени (конвективным $\tau_k = \L/V_0$ и акустическим $\tau_a  = L/a$), 
 сопряжено с  вычислительными трудностями \cite{roach,lapin,zhm01}.
	В явных разностных схемах временной шаг ограничен величиной порядка $h/(V + a) \approx h/a \approx \tau_a/N$ и они эффективны лишь для моделирования собственно акустических явлений \cite{fed}. Неявные схемы безусловно устойчивы (для линеаризованных уравнений), однако при стремлении числа Маха к нулю возрастает трудоемкость расчетов и падает их точность 
\cite{roach,lapin}. Это связано с малостью относительных изменений давления в гипозвуковом потоке и проявляется особенно остро, если в качестве зависимой переменной используется плотность. Записывая уравнение состояния в виде $p = (a^2/\gamma) \rho$ ($\gamma$ - показатель адиабаты), легко видеть, что малые изменения плотности приводят к большим изменениям градиента давления в уравнении движения 
$$\frac{1}{\rho_0 V_0^2} \nabla p \approx \frac{1}{\gamma M^2} \nabla \left( \frac{\rho}{\rho_0} \right) \gg \nabla \left( \frac{\rho}{\rho_0} \right).$$
 Наконец, последняя по счету, но не по значимости, проблема --- постановка условий на проницаемых границах 	\cite{fed}.
	
	Эти трудности не возникают при использовании гипозвуковых уравнений Навье-Стокса  \cite{40}, следующих из 
полных уравнений Навье-Стокса в предположении, что \cite{41}:
{число Маха мало $M^2\ll 1$;} 
{параметр гидростатической сжимаемости $\varepsilon={gL}/{R_g T_0}$ мал $ \varepsilon \ll 1$ ( {\itмелкая} конвекция \cite{zey});}
{характерное время велико в сравнении с акустическим, что позволяет выбрать $L/V_0$ масштабом времени (число Струхаля $Sh = 1$.}).  Приближение Буссинеска требуют дополнительно малости изменений плотности $\Delta \rho/ \rho_0 \ll 1$.

Приведем вывод  на примере  однородного  совершенного  газа. В  общепринятых обозначениях уравнения Навье-Стокса имют вид
\begin{eqnarray} 
\frac{d \rho}{d t} + \rho \nabla \cdot {\bf V}  = 0, \\
\rho \frac{d {\bf V}}{d t}  = - \frac{1}{\gamma M^2} \nabla p + \frac{1}{Fr} \rho {\bf j} + \frac{1}{Re} \left[ 2 Div (\mu \dot{S})  - \frac{2}{3} \nabla (\mu \nabla \cdot {\bf V}) \right],\\
\rho\frac{d T}{d t }  =  \frac{\gamma - 1}{\gamma} \frac{d p}{d t} + \frac{1}{Re Pr} \nabla \cdot (\lambda \nabla T) + \frac{1}{Re}(\gamma - 1) M^2 \left[2 \mu \dot{S}^2 - \frac{2}{3} (\nabla \cdot {\bf V})^2  \right],\\
\rho T  = p.
\end{eqnarray} 
В разложении по двум малым параметрам ($ \delta = \gamma  M^2$ и $\epsilon$)
$$f = f^{(0)} + \delta f^{(10)} + \epsilon f^{(01)} + \cdots$$
первое нетривиальное приближение содержит $\rho^{(0)}, {\bf V}^{(0)}, T^{(0)}$ (ниже индекс опущен), $p^{(0)}, p^{(01)}, p^{(10)}$: 
\begin{eqnarray} 
\nabla p^{(0)} = 0,\\
\frac{d \rho}{d t} + \rho \nabla \cdot {\bf V}  = 0, \\
\rho \frac{d {\bf V}}{d t}  = -  \nabla p^{(1)} + \frac{1}{Fr} \rho {\bf j} + \frac{1}{Re} \left[ 2 Div (\mu \dot{S})  - \frac{2}{3} \nabla (\mu \nabla \cdot {\bf V}) \right],\\
\rho\frac{d T}{d t }  =  \frac{\gamma - 1}{\gamma} \frac{d p^{(0)}}{d t} +\frac{1}{Re Pr} \nabla \cdot (\lambda \nabla T) \\
\rho T  = p^{(0)}.
\end{eqnarray} 

\noindent Очевидно, масштаб избыточного (динамического) давления $p^{(1)}$ равен $\rho_0 V_0^2$, а $p^{(0)}$ может зависить только от времени. Член $\nabla p^{(1)}$ есть сумма двух членов одного порядка   $\nabla p^{(10)}$  и $(\epsilon/\delta)\nabla p^{(01)}$

Гипозвуковые уравнения подобно несжимаемым уравнениям Навье-Стокса не описывают акустические явления: сжимаемость проявляется только в локальном тепловом расширении. Соотношение, обобщающее условие несжимаемости на случай 
гипозвуковых уравнений,  имеет вид  \cite{41,zhm90_m}
\begin{equation}
\nabla \cdot {\bf V} = \frac{1}{p^{(0)}} \left[\frac{1}{Re Pr} \nabla \cdot (\lambda \nabla T)  - \frac{1}{\gamma} \frac{d p^{(0)}}{d t}  \right].
\end{equation}
Заметим, что подобное соотношение для частного случая получил еще Обербек \cite{ober} --- из уравнений (7) и (19) его работы следует стационарный вариант обобщенного условия несжимаемости.

Для смеси  $N_s$ компонент консервативная форма  стационарных гипозвуковых уравнений  в  размерном виде  имеет вид  
\begin{eqnarray} 
\nabla \cdot \rho {\bf V} =0\\ 
\nabla \cdot \left( \rho {\bf VV} +p\hat I-\hat \tau \right) -\left( \rho 
-\rho _0\right) {\bf g}=0\\ 
\nabla \cdot \left( \rho {\bf V} h+{\bf q} \right) =0\\ 
\nabla \cdot \left( \rho {\bf V} c_s+{\bf J}_s\right) =W_s\,,\;s=1,2,\ldots 
,N_s 
\end{eqnarray} 
где $\rho$ --- плотность смеси, 
${\bf V}$ --- среднемассовая скорость, 
$h$ --- удельная энтальпия, 
$c_s$ --- массовая доля s-й компененты, 
$p$  --- динамическое давление, 
$\tau$  --- тензор вязких напряжений, ${\bf q}$ --- тепловой поток, ${\bf J_s}$ 
- диффузионный поток s-й компененты. 
Для замыкания системы требуются уравнение состояния смеси совершенных газов: 
\begin{equation} 
\rho {R\over m}T=p_0=const\,,\; 
1/m=\sum\limits_{s=1}^{N_s}\left( c_s/m_s\right)\\ 
\end{equation} 
и определения молекулярных потоков: тензора вязких напряжений  
\begin{equation}
\hat \tau =-{2\over 3}\mu \left( \nabla \cdot {\bf V}\right) \hat I+\mu 
\left( \nabla {\bf V} +{\bf V} \nabla \right),
\end{equation}
теплового потока
\begin{equation}
{\bf q} =-\lambda \nabla T+\sum\limits_{s=1}^{N_s}h_s{\bf 
J}_s+p_0\sum\limits_{s=1}^{N_s}k_s^T{\bf J}_s/\left( \rho c_s\right)
\end{equation}
и диффузионного потока  $s$й компоненты
\begin{equation}
{\bf J}_s=-\rho D_s\left( \nabla c_s+{m_s\over m}k_s^T{\nabla T\over 
T}\right). 
\end{equation} 

В случае разбавленных смесей газодинамическая задача отщепляется от задачи массопереноса 
\cite{42,43}. 
При высокой концентрации активных реагентов в потоке несущего газа возможна нуклеация кластеров --- модель, основанная на уравнениях для моментов фунции распределения, описана в  \cite{vor,44}.


\subsubsection{Радиационный теплообмен (РТ)}
Теплообмен излучением --- важный (часто  основной) механизм при росте кристаллов. Его используют и для нагрева в варианте метода зонной плавки \cite{45}. 

Различают три случая взаимодействия излучения со средой ($\alpha$ --- коффициент поглощения, $L_0$ --- характерный размер): $\alpha L_0 \gg 1$; $\alpha L_0 \sim 1$; $\alpha L_0 \ll 1$. 
В первом --- непрозрачная среда --- тепло переносится молекулярной  теплопроводностью.
Полупрозрачная среда приближенно описывается эффективной теплопроводностью 
$\lambda_{eff} = \lambda + \lambda_r$, где радиационная теплопроводность  $\lambda_r = 16 n^2 \sigma T^4 / 3 \alpha$, $n$ --- коэффициент преломления,   $\sigma$  --- постоянная Стефана-Больцмана.

В простейшем случае --- все объекты прозрачные или непрозрачные, а все поверхности отражают диффузно ---  достаточно использовать конфигурационные факторы (view factors) $F_{ij}$ \cite{47,ern}. 
Расчет этих величин с учетом затенения подробно описан в \cite{47,vf1}. Полный поток излучения на $i$-й поверхностный элемент 
($i=\overline{1,N_e}$, где $N_e$ --- число поверхностных элементов)
\begin{equation} 
q_i^{in}=\sum\limits_{j=1}^{N_e}{q_j^{out}F_{ij}}, 
\end{equation} 
где  $q^{in}$ и $q^{out}$ - потоки излучения к и от поверхности.

Задача усложняется, если надо учесть зеркальное отражение и/или
полупрозрачность кристалла \cite{z10,ruk,z8} (многие оксиды, галиды и флюориды --- лазерные, оптические, ювелирные кристаллы --- полупрозрачны), спектральные зависимости оптических свойств (в самой простой двухполосной модели среда предполагается либо полностью прозрачной, либо непрозрачной  в соответствующих диапазонах длин волн \cite{47}).

\subparagraph{Полупрозрачность}
Коэффициент поглощения определяет поглощение и эмиссию излучения в объеме, коэффициент преломления --- отражение и преломление на границах кристалла.

РТ через кристалл существенен для отвода тепла от фазовой границы; РТ в кристалле может вести  к изменению формы интерфейса \cite{48,49} и к неустойчивости фронта кристаллизации \cite{m2:yuf87}.
Распределение потока излучения от фронта кристаллизации может кардинально меняться в процессе роста с увеличением длины кристалла с большим показателем преломления из-за эффекта полного внутреннего отражения 
\cite{50}. Встраивание микропузырьков в кристалл может привести к немонотонному распределению температуры в оптических кристаллах  \cite{51}. 
РТ в расплаве обычно менее важен в силу большого коэффициента поглощения, однако пренебрежение внутренним РТ в расплаве оксида 
LiNbO$_3$  приводит к фиктивной структуре спиц \cite{52}. 
Полупрозрачные свойства инкапсулянта B$_2$O$_3$ 
рассматривались в работах  \cite{47,53}.

Различные приближения используются в анализе РТ при росте кристаллов.
Сопоставление с экспериментальными данными 
\cite{54} показывает, что учет излучения с помощью 
радиационной теплопроводности  $\lambda_r$ (например, для изучения роста 
CaF$_2$ из расплава \cite{55}) не позволяет корректно описать распределение температуры в системе, поэтому нужны более изощренные подходы ---
метод дискретных обменных факторов \cite{56} или метод характеристик 
\cite{57}.

Авторы \cite{49,m2:y3} изучали РТ в кристаллах иттрий-алюминиевого и гадолиний-галиевого гранатов в предположении, что кристалл прозрачен, а его поверхность покрыта исчезающе тонкой непрозрачной пленкой, что позволило пренебречь преломлением. Утверждается, что пленка --- результат паразитного осаждения примесей в реальном росте
(что не мешает авторам указывать учет преломления как будущую работу).
Заметим, что депозиты на стенках {\em эпитаксиального} реактора усложняют расчет РТ, если их толщина сравнима с длиной волны излучения, однако это явление не столь распространено при росте монокристаллов из расплава.

Тсукада и др. \cite{m2:y13}, считая коэффициент поглощения достаточно большим, использовали для РТ Р1-прилижение; подобный подход применен в 
 \cite{m2:y4,m2:y15,m2:y16} к изучению влияния коэффициента поглощения на фронт кристаллизации ниобата лития и гранатов.

Обобщенным методом дискретных обменных факторов \cite{56} было исследовано влияние условия на верхней границе на распределение температуры, форму фазовой границы и устойчивость роста иттрий-алюминиевого граната из расплава  \cite{m2:y15}, однако поверхность кристалла предполагалась непрозрачной. Теплообмен при росте силленита  и условия формирования плоского фронта кристаллизации  исследовались в работах  \cite{47,53,m2:y16}.

Галазка и др.  \cite{galaz} использовали многополосную модель излучения,
основанную на измерениях в спектральном диапазоне  
0.8 $\mu$m - 4.2 $\mu$m с шагом 50 nm, для исследования роста иттрий-алюминиевого граната методом Чохральского. Авторы вычислили средние коэффициенты поглощения Планка $\alpha_P$ и Росселанда $\alpha_R$ и тем самым оптическую толщину $\tau = L \sqrt{\alpha_P \alpha_R}$ и фактор отклонения от "`серой"' модели излучения 
$\eta = L \sqrt{\alpha_P / \alpha_R}$.  

Авторы  \cite{ban} моделировали тот же процесс в осесимметричной постановке в единой области, содержащей твердую, жидкую и газовую фазу с своими материальными параметрами; излучение описывалось  $S_N$ методом дискретных ординат. Обнаруженный перегрев кристалла выше точки плавления, очевидно, является вычислительным артефактом. В отличие от {\em переохлажденной} жидкости как метастабильного состояния (со значительным отклонением от температуры кристаллизации - например, почти 40 K для воды \cite{z9} и 20 K 
для германата висмута \cite{74}),   {\em перегретая} твердая фаза существовать не может, поскольку для плавления нет энергетического 
барьера\footnote{Неравновесное состояние твердого тела с температурой выше температуры плавления может быть достигнуто путем ультрабыстрого  нагрева --- интенсивным лазерным излучением или с помощью ударных волн 
\cite{shock,shock1}. }.
При температурах, близких (снизу) к температуре кристаллизации наблюдается "`предплавление"' (premelting) - появление тонкого жидкого слоя на твердой поверхности - для широкого класса материалов, включая воду, металлы, полупроводники  \cite{pre}.

Значительно меньшее внимание уделяется зеркальному отражению от поверхности кристалла (зеркальное отражение от поверхности {\em расплава} кремния в методе Чохральского оказывает некоторое влияние на распределение температуры в окрестности мениска \cite{m2:czsi}). Первые расчеты проведены методом характеристик для роста германата висмута методом Чохральского с низкими температурными градиентами \cite{57}, при этом форма фронта кристаллизации считалась неизменной; о самосогласованных расчетах сообщалось в работе  \cite{50}. 
Показано, что изменение формы фронта в процессе роста определяется зеркальным отражением от внутренней поверхности кристалла, прежде всего от ее конической части.

Ниже описана постановка локальной задачи о переносе излучения в полупрозрачном кристалле и зазоре между кристаллом и тиглем в осесимметричном случае
\cite{z8,50,57}; трехмерный вариант рассмотрен в работах \cite{m2:mam,m2:y_m}.
Расплав предполагается непрозрачным, учитывается зависимость коэффициента поглощения кристалла от длины волны. Стенка тигля отражает диффузно, а  поверхность кристалла ---  диффузно или зеркально.
Сопряжение локальной и глобальной задач обеспечивается следующими условиями:
\noindent
\textit{свободная поверхность расплава} ($z = 0$):
\begin{equation}
- \lambda_L \nabla T_L \cdot \vec{n} = \epsilon_L (\sigma T_L^4 - q^{inc});
\end{equation}
\noindent
\textit{граница кристалл/расплав} ($z = h (r,t)$):
\begin{equation}
T_L = T_S = T_m;
\end{equation}
\begin{equation}
-\lambda_S \vec{n} \cdot \nabla T_S + \lambda_L \vec{n} \cdot \nabla T_L + q_{rad} = L (\vec{n} \cdot \vec{e_z}) (V - \dfrac{dh}{dt}),
\end{equation}
где $\vec{n}$ --- единичная нормаль в сторону области кристалла, $\vec{e_z}$
---  орт, $L$ --- удельная теплота кристаллизации, $V$ --- скорость вытягивания, $T_m$ --- температура плавления, индексы L и S  соответствуют жидкой и твердой фазам, $\epsilon_L$ --- коэффициент черноты расплава, $q^{inc}$ и $q_{rad}$ --- плотности потока излучения на свободную поверхность расплава и на границу кристалл/расплав,  $h$ --- координата фронта кристаллизации. В предположении стационарности тройной точки скорость вытягивания определяется соотношением  
\begin{equation}
L ( \vec{n} \cdot \vec{e_z}) V = - \lambda_S \vec{n} \cdot \nabla T_S +
\lambda_L  \vec{n} \cdot \nabla T_L + q_{rad} \rvert_{r = R},
\end{equation}
\noindent где  $R$ --- радиус кристалла и для фронта кристаллизации получаем 
\begin{equation}
\dfrac{\lambda_S \vec{n} \cdot \nabla T_S - \lambda_L \vec{n} \cdot \nabla T_L - q_{rad}} {(\vec{n} \cdot \vec{e_z})} -
\dfrac{\lambda_S \vec{n} \cdot \nabla T_S - \lambda_L \vec{n} \cdot \nabla T_L - q_{rad}} {(\vec{n} \cdot \vec{e_z})}
\rvert_{r = R_0} = L \dfrac{dh}{dt}
\end{equation}
В стационарной задаче время $t$ играет роль итерационного параметра.
Уравнение энергии для кристалла имеет вид:
\begin{equation}
\nabla \cdot (\lambda_S \nabla T_S) = \nabla \cdot \int Q^{rad}(\vec{r}, \lambda) d \lambda =
\int \limits_{\lambda}^{} \int \limits_{\Omega = 4 \pi}^{} k_{\lambda}
(B_{\lambda}(T_S) - I_{\lambda}(\vec{r},\Omega)) d \lambda d \Omega
\end{equation}
\noindent где  $Q^{rad}$ --- плотность потока излучения, $\lambda$ --- длина волны излучения,
$I_{\lambda}$ --- интенсивность излучения, $k_{\lambda}$ --- коэффициент поглощения и
$B_{\lambda}(T)$ --- интенсивность излучения черного тела.
Конвективным членом можно пренебречь в силу малости скорости вытягивания.
Граничные условия ставятся следуюшим образом:

\noindent
\textit{боковая поверхность кристалла}:
\begin{equation}
-\lambda_S \dfrac{\partial T_S}{\partial N} = \epsilon_S W_1(T) (n^2 \sigma {T_L}^4 - {q_1}^{inc}  )
\end{equation}
\noindent где $N$ --- внешняя нормаль к боковой поверхности, $n$ и $\epsilon_S$ --- коэффициенты преломления и черноты, $q_1^{inc}$ --- падающий поток излучения, $W_1(T)$ --- доля излучения черного тела в диапазоне длин волн, в котором кристалл непрозрачен 
$
W_1(T) = {\int \limits_{\Delta_{1 \lambda}}^{}B_{\lambda}(T) d \lambda }/(n^2 \sigma T^4 / \pi)
$.

\noindent
\textit{ось симметрии}:
$
\dfrac{\partial T_S}{\partial r} = 0.
$

Интенсивность излучения $I_{\lambda}$ описывается уравнением переноса излучения, которое в пренебрежении рассеянием имеет вид 
\begin{equation}
\dfrac{\sin \theta}{r} \left[ {\cos \phi} \dfrac{\partial(rI_{\lambda})}{\partial r} -
\dfrac{\partial({\sin \phi} I_{\lambda})}{\partial \phi} \right] + {\cos \theta} \dfrac{\partial I_{\lambda}}{\partial z} +
k_{\lambda} I_{\lambda} = k_{\lambda} B_{\lambda}
\end{equation}
где $\phi = \phi_{\Omega} - \phi_r$, $\phi_r$, $r$, $z$ цилиндрические координаты,  $\phi_{\Omega}$, $\theta$  --- сферические координаты, определяющие направление луча
$\vec{\Omega} = (\sin \theta \cos \phi_{\Omega}, \sin \theta \sin \phi_{\Omega}, \cos \theta)$.

Решение возможно при выборе нулевым  $\phi_{\Omega}$ или $\phi_r$.  Детальное описание первого подхода и численной реализации можно найти в работах \cite{49,57,m2:y22}. Идея метода в следующем.

Для $\phi_{\Omega} = 0$ решение ищется в точках $(r, z, \phi_r)$ 
трехмерного пространства для лучей, параллельных плоскости $rz$, т. е. для направлений $\vec{\Omega} =(\sin \theta,0,\cos \theta)$.  
Тогда преобразование к декартовым координатам  $x,y,z$ где $x = r \cos \phi_r, y = r \sin \phi_r$ дает:
\begin{equation}
\sin \theta \dfrac{\partial I_{\lambda}}{\partial x} +
\cos \theta \dfrac{\partial I_{\lambda}}{\partial z} +
k_{\lambda} I_{\lambda} = k_{\lambda} B_{\lambda}(T)
\end{equation}

Граничные условия определяются типом границы:

\textit{Непрозрачная диффузно отражающая (стенка тигля)}:
\begin{equation}
I_{\lambda}^{out}(\vec{r}, \theta) = \rho_d \dfrac{q_{\lambda}^{inc}(\vec{r}, z)}{\pi} +
(1 - \rho_d) B_{\lambda}(T_{cr}), \qquad \qquad \vec{\Omega} \cdot \vec{\nu} > 0
\end{equation}
\begin{equation*}
q_{\lambda}^{inc}(r, z) = 2 \int \limits_{\vec{\Omega}^{\prime} \cdot \vec{\nu}^{\prime} < 0}
|\vec{\Omega}^{\prime} \cdot \vec{\nu}^{\prime}| I_{\lambda} (r^{\prime}, \theta^{\prime})
\sin \theta^{\prime} d \theta^{\prime} d \phi^{\prime}\\
\end{equation*}
\noindent
где ${\nu}$ --- внутрення нормаль, $\rho_d$ --- диффузный коэффициент отражения, $q_{\lambda}^{inc}$ --- падающий поток излучения,
$\vec{\nu^{\prime}} = (\sin \theta_{\nu} \cos \phi^{\prime}, \sin \theta_{\nu} \sin \phi^{\prime}, \cos \theta_{\nu})$, $\phi^{\prime} \in (0, \pi)$  и индекс "out" означает излучение, уходящее от границы. Здесь и ниже
$\vec{r} = (r \cos \phi_r, r \sin \phi_r, z)$
and ${\vec{r}}^{\prime} = (r \cos \phi_r^{\prime}, r \sin \phi_r^{\prime}, z)$;
углы  $\phi_r$ и  $\phi_r^{\prime}$ в (7a) определяют положение граничных точек.

\textit{Непрозрачная зеркально отражающая (свободная поверхность)}:
\begin{equation}
I_{\lambda}^{out}(\vec{r}, \theta) = \rho_s I_{\lambda} ({\vec{r}}^{\prime}, \theta^{\prime}) +
(1 - \rho_s) B_{\lambda}(T_L), \qquad \qquad \vec{\Omega} \cdot \vec{\nu} > 0
\end{equation}

\noindent
где $\rho_s$ - спектральный коэффициент отражения, а $\theta^{\prime}$ - полярный угол направления $\vec{\Omega}^{\prime}$ луча, падающего на границу и отраженного в направлении $\vec{\Omega}$ с полярным углом $\theta$. Связь
$\vec{r},\theta$  и  ${\vec{r}}^{\prime},\theta^{\prime}$ дается законом отражения 
${\vec{\Omega}}^{\prime} = \vec{\Omega} -2 (\vec{\Omega} \cdot \vec{\nu}) \vec{\nu}$.

\textit{Прозрачная дифффузно отражающая (боковая поверхность)}:
\begin{equation}
I_{\lambda_1}^{out}(\vec{r}, \theta) = \rho_{d_1} \dfrac{q_{\lambda_1}^{inc}(\vec{r}, z)}{\pi} +
(1 - \rho_{d_2})\dfrac{q_{\lambda_2}^{inc}(\vec{r}, z)}{\pi} , \qquad \qquad \vec{\Omega} \cdot \vec{\nu} > 0
\end{equation}
\begin{equation*}
I_{\lambda_2}^{out}(\vec{r}, \theta) = \rho_{d_2} \dfrac{q_{\lambda_2}^{inc}(\vec{r}, z)}{\pi} +
(1 - \rho_{d_1})\dfrac{q_{\lambda_1}^{inc}(\vec{r}, z)}{\pi} , \qquad \qquad \vec{\Omega} \cdot \vec{\nu} < 0
\end{equation*}
\noindent
где индексы 1 и 2 обозначают газ и кристалл, соответственно,  $\rho_{d_1}$ и  $\rho_{d_2}$ связаны как
\begin{equation}
1 - \rho_{d_1} = (1 - \rho_{d_2}) n^2
\end{equation}

\textit{Прозрачная зеркально отражающая (боковая поверхность)}:
\begin{equation}
I_{\lambda_1}^{out}(\vec{r}, \theta) =
\rho_{s_1} I_{\lambda_1}^{inc}(\vec{r^{\prime}}, \theta^{\prime}) +
(1 - \rho_{s_1})\dfrac{I_{\lambda_2}^{inc}({\vec{r}}^{\prime \prime}, \theta^{\prime
		\prime})}{n^2} , \qquad \qquad \qquad \vec{\Omega} \cdot \vec{\nu} > 0
\end{equation}
\begin{equation*}
\dfrac{I_{\lambda_2}^{out}(\vec{r}, \theta)}{n^2} =
\rho_{s_2} \dfrac{I_{\lambda_2}^{inc}({\vec{r}}^{\prime}, \theta^{\prime})}{n^2}
(1 - \rho_{s_2}) I_{\lambda_1}^{inc}(\vec{r^{\prime \prime}}, \theta^{\prime \prime})
, \qquad \qquad \vec{\Omega} \cdot \vec{\nu} < 0
\end{equation*}
\noindent
где $\theta^{\prime \prime}$ --- полярный угол направления ${\vec{\Omega}}^{\prime \prime}$ луча, падающего на границу и отраженного в направлении   $\vec{\Omega}$ с полярным углом $\theta$. Направления 
${\vec{\Omega}}^{\prime \prime}$ и $\vec{\Omega}$
связаны законом Снелла, а ${\vec{r}}^{\prime \prime}$ и $\theta^{\prime
	\prime}$ находятся по  $\vec{r}$ и  $\theta$.
Коэффициенты зеркального отражения определяются формулами Френеля. Заметим, что $\rho_{s_2} = 0$, если $|\vec{\Omega} \cdot \vec{\nu}| > \cos \phi_B$, где $\phi_B$ --- критический угол полного внутреннего отражения. 

Для решения уравнения переноса излучения используется метод дискретного переноса, адаптированный к осесимметричному случаю. Метод интегрирует уравнение вдоль выбранных направлений 
 $\vec{\Omega_j} = (\sin \theta_j, 0, \cos \theta_j), \theta_j \in (0, \pi) ,
 j = 1, ..., N_{\theta}$.
Решение ищется в области
 $\mathcal{D}_{r \phi z} = \mathcal{D}_{rz} \bigotimes [0,\pi]$  в  переменных  $r, \phi, z$
где $\mathcal{D}_{rz}$ --- меридиональное сечение области.
 Область  $\mathcal{D}_{r \phi z}$  разбивается на 3D ячейки (для различных направлений разбиение может быть разным)  с использованием разбиения плоской области.  3D ячейки имеют форму
 $V_n = \{ \vec{r}: (r, z) \in t_m, \quad \phi \in (\phi_{i-1,j}, \phi_{i,j})  \} $  ,
 где $t_m$ - планарная ячейка,
 $\phi_{i,j} = \pi i/N_{\phi,j}, i=1, ..., N_{\phi,j}$ ($N_{\phi,j}$ может зависить от направления).
Интенсивность излучения дается средними значениями на гранях ячеек, лежащих на границах блоков. Луч испускается из центра граничной ячейки  и  прослеживается до падения на другую границу. 
Интенсивность излучения известна из граничных условий или с предыдущей итерации.
Отклонением точки падения от центра грани пренебрегается. Уравнение переноса излучения интегрируется аналитически от точки падения до точки испускания.
Процедура повторяется до достижения сходимости.

Падающий поток на границе аппроксимируется как 
 \begin{equation}
 q^{inc} = \sum_{i=1}^{N_{\theta}}w_i
 \sum_{m \in \Gamma_k, \qquad
 	\vec{\Omega}_i \cdot \vec{n}_m > 0} S_{mi}I_{mi}^{inc}
 \end{equation}
 \noindent
 где  $\Gamma_k$ --- множество граней, которые могут быть совмещены с гранью  $k$ вращением вокруг оси симметрии, $\vec{n}_m$  --- "`средняя"' внешняя нормаль к грани $m$,
 $S_{mi} = | \vec{\Omega}_i \cdot \vec{n}_m| S_m$   и $S_m$ --- площадь грани  $m$.
 Специальный алгоритм был разработан для вычисления дивергенции вектора потока излучения внутри ячейки \cite{m2:y22}.
Метод был применен к моделированию роста кристаллов германата висмута, сапфира, алюминиевых гранатов  \cite{z10,50,z8,ruk,m2:y_m}.

Универсальным методом расчета распространения излучения
с учетом  поглощения,  рассеяния, преломления, диффузного и зеркального отражения  является метод Монте-Карло 
\cite{how, ker}.
Общий подход (метод трассировки лучей), использующий обобщенные конфигурационные факторы для поверхностных и полупрозрачных объемных элементов в терминах диффузных  и абсорбционных угловых коэффициентов, предложен в  \cite{58} для решения уравнения баланса излучения
в направлении   $\hat{\eta}$ 
\begin{multline} \label{eq:base}
\frac{d I_\lambda (\vec{r}, \eta)} {d S} = - (k_\lambda +
\sigma_{\eta, \lambda}) I_\lambda (\vec{r}, \eta) + k_\lambda
I_{b, \lambda}(T) +\\ + \frac{\sigma_{\eta, \lambda}}{4 \pi}
\int_{4 \pi} I_\lambda (\vec{r}, \eta') P_\lambda (\eta' \to \eta)
d \omega
\end{multline}
где $k_\lambda$ и $\sigma_{\eta, \lambda}$ - коэффициенты отражения и рассеяния,  $P_\lambda(\eta'\to\eta)$ - фазовая функция от направления 
$\eta'$ к $\eta$.

Метод свободен от т. н. "`лучевого"' эффекта, свойственного методам, использующим дискретный набор направлений излучения, однако  модель эмиссии лучей, описанная в \cite{58}, неприменима к средам с большим коэффициентом поглощения и, как следствие, к многополосному моделированию излучения. 
Модель эмиссии, свободная от этого недостатка, предложена в  \cite{59,koch1}.
Рассматривается излучательный элемент (выпуклый многогранник или многоугольник), среда считается изотропной при постоянной температуре.
Средняя толщина элемента в направлении $\eta$ определяется как 
\begin{equation} \label{eq:av_thick}
\bar{S}\equiv \frac{V}{A(\eta)},
\end{equation}
где $V$ и $A(\eta)$ - объем и площадь проекции элемента на плоскость, ортогональную к $\eta$.

Для единого описания поверхностей и активной среды диффузная отражательная способность поверхности и альбедо среды переопрелеляютя как 
$\Omega^D$ и вводится зеркальная отражательная способность поверхности $\Omega^S$.
Тогда энергия, излученная элементом, 
\begin{equation}\label{eq:general}
d Q_{J,i,\lambda}(\eta) =\\ A_i(\eta)\left[ \left( 1 - \Omega^D -
\Omega^S \right) I_{b,\lambda} + \Omega^D I_{\lambda}^D\right]
\times  \left[1 - \exp(-\beta_\lambda \bar{S_i}) \right]d
\omega,
\end{equation}
где $\beta_\lambda = k_\lambda + \sigma_{\eta, \lambda}$ - коэффициент экстинкции в направлении $\eta$ для длины волны  $\lambda$, 
$\beta_\lambda \bar{S_i}\gg 1$ для поверхностного элемента и 
$\Omega^S = 0$ для объемного.

Интегрирование по телесным углам дает 
\begin{equation}\label{eq:int_gen}
Q_{J,i,\lambda} = \left[ \left( 1 - \Omega^D - \Omega^S
\right)I_{b,\lambda} + \Omega^D I_{\lambda}^D\right]  \\ \times
\int_{4 \pi}A_i(\eta)\left[1 - \exp(-\beta_\lambda \bar{S_i})
\right]d \omega
\end{equation}

Согласно \cite{58} вводится понятие излучающей поверхности для элемента, который может быть поверхностным или объемным 
\begin{equation}\label{eq:eff_area}
A_i^R \equiv \frac{1}{\pi} \int_{4 \pi}A_i(\eta)\left[1 -
\exp(-\beta_\lambda \bar{S_i}) \right]d \omega,
\end{equation}
где $A_i^R$ идентично площади элемента, у которого не рассматривается обратная сторона.
Наконец, поток энергии, которая излучена и изотропно рассеяна элементом,  
\begin{equation}\label{eq:gen_form}
Q_{J,i,\lambda} = \pi (\epsilon_i I_{b, \lambda} + \Omega^D
I_\lambda^D)A_i^R,
\end{equation}
где $\epsilon_i = 1 - \Omega^D_i - \Omega^S_i$ и 
$Q_{J,i,\lambda}$ --- интенсивность диффузного излучения.

\subparagraph{Вычисление конфигурационных факторов}
Угловой коэффициент экстинкции $F_{i,j}^E$ --- доля энергии, излученной элементом $i$, которая поглощена, изотропно рассеяна или диффузно отражена элементом $j$
\begin{multline}\label{eq:extVF}
F_{i,j}^E = \int_{4 \pi} f_i^j(\eta) A_i(\eta)\left[ 1 -
\exp(-\beta_{\lambda, i} \bar{S_i})\right] \\ \times \left[1 -
\exp(-\beta_{\lambda, j} S_j) \right] d \omega /\left( \pi A_i^R
\right),
\end{multline}
где $S_j$ - длина пути луча в объемном элементе $j$, а $f_i^j(\eta)$ - доля энергии, излученной элементом  $i$ в направлении  $\eta$, которая достигла элемента $j$.
Абсорбционный и диффузионный угловые коэффициенты определены как
\begin{equation}\label{eq:FA_FD}
F_{i,j}^A = \frac{\epsilon_j F_{i,j}^E}{1 - \Omega_j^S}, \qquad
F_{i,j}^D = \frac{\Omega_j^D F_{i,j}^E}{1 - \Omega_j^S}.
\end{equation}

Вычисление этих коэффициентов проводится в процессе трассировки испущенного луча через систему объемных и поверхностных элементов \cite{59,koch1}. При попадании луча в поверхностный элемент $j$  он расщепляется на поглощенный, рассеянный и отраженный. При взаимодействии луча с поверхностью объемного полупрозрачного  элемента он разделяется в соответствии с соотношениями Френеля на два: отраженный и преломленный. 
Поглощенная доля энергии дает вклад в $F_{i,j}^A$, рассеянная (на поверхности элемента и в объеме) в   $F_{i,j}^D$.
Путь луча прослеживается, пока интенсивность не уменьшится до заданого значения.

Вводя излучательную мощность элемента $i$ в диапазоне $\Lambda$  
\begin{equation}\label{eq:Qtau}
Q_{\tau,i,\Lambda} \equiv A_{i, \Lambda}^R \epsilon_{i,\Lambda}
n_\Lambda^2 \sigma \gamma_\Lambda(T) T_i^4,
\end{equation}
для суммарного потока в системе $N$ поверхностных и объемных элементов имеем 
\begin{equation}\label{eq:net_Geph}
Q_{i,\Lambda}^{net} = \sum_{j=1}^N F^G_{i,j,
\Lambda}Q_{\tau,j,\Lambda},
\end{equation}
где $F^G_{i,j, \Lambda}$ --- элементы матрицы Гебхарда $\mathbf{F}^G_\Lambda$ ($\mathbf{E}$ --- единичная матрица):
\begin{equation}\label{eq:Gephard}
\mathbf{F}^G_\Lambda = \mathbf{E} -
\left(\mathbf{F}^A_\Lambda\right)^T\left(\mathbf{E} -
\left(\mathbf{F}^D_\Lambda\right)^T\right)^{-1}.
\end{equation}
Полный поток излучения находится суммированием по всем спектральным полосам  \cite{59,koch1}
\begin{equation}\label{eq:multirange}
Q_{i}^{net} = \sum_\Lambda Q_{i,\Lambda}^{net} =  \sum_\Lambda
\sum_{j = 1}^N F^G_{i,j, \Lambda}Q_{\tau,j,\Lambda},
\end{equation}
и окончательно
\begin{equation}\label{eq:ML_reWri}
\vec{Q}^{net} = \left(\sum_\Lambda \mathbf{F}^G_\Lambda
\mathbf{A}^R_\Lambda \mathbf{\epsilon}_\Lambda
\mathbf{n}^2_\Lambda \right) \sigma \vec{T}^4,
\end{equation}
где матрицы $\mathbf{A}^R_\Lambda, \mathbf{\epsilon}_\Lambda, \mathbf{n}_\Lambda$ - диагональные.

В осесимметричном случае трассировка луча по-прежнему имеет трехмерный характер, но метод может быть существенно упрощен и ускорен. Его нельзя непосредственно применить к тороидальным элементам треугольного сечения, т. к. эффективная поверхность определена для выпуклого элемента. Поэтому каждый элемент разбивается на $N^{seg}$ равных многогранников или многоугольников (для объемных и поверхностных элементов соответственно).
Введем индекс $l$ для сегментов $i$-го элемента и индекс $k$ для  $j$-го. Очевидно, $F_{i,l,j,k} = F_{i,l+1,j,k+1}$
и коэффициенты в кольцеобразном множестве получаются по значениям для первого сегмента циклической перестановкой индексов \cite{zhm01},
а для $i$-го элемента
\begin{equation}\label{eq:net_Geph_seg}
Q_{i}^{net} = \sum_{l=1}^{N_{seg}} \sum_{j=1}^{N}
\sum_{k=1}^{N_{seg}} F^G_{i,l,j,k }Q_{\tau,j,k}.
\end{equation}

Матрица размерности  $(N \cdot N_{seg} \times N\cdot N_{seg})$ сводится  к матрице  размерности $N \times N$ использованием идентичности эффективной поверхности  $A^R_{j,k}$, температуры $T_{j,k}$ и степени черноты 
$\epsilon_{j,k}$ для любого $k$ при фиксированном  $j$ 
\begin{multline}\label{eq:net_rewr}
Q_{i}^{net} = \sum_{l=1}^{N_{seg}} \sum_{j=1}^{N}
\sum_{k=1}^{N_{seg}} F^G_{i,l,j,k}A_{j,k}^R \epsilon_{j,k} n^2
\sigma \gamma(T_{j,k}) T_{j,k}^4 = \\ \sum_{j=1}^{N} \left[
\sum_{l=1}^{N_{seg}} \sum_{k=1}^{N_{seg}}F^G_{i,l,j,k}
\right]A_{j,1}^R \epsilon_{j,1} n^2 \sigma \gamma(T_{j,1})
T_{j,1}^4 = \\ \sum_{j=1}^{N} \left[ N_{seg}
\sum_{l=1}^{N_{seg}}F^G_{i,l,j,1}\right] A_{j,1}^R \epsilon_{j,1}
n^2 \sigma \gamma(T_{j,1}) T_{j,1}^4 =
\sum_{j=1}^{N}\tilde{F}^G_{i,j} \tilde{Q}_{\tau,j}.
\end{multline}

Модифицированная матрица Гебхарда $\tilde{\mathbf{F}}^G$ 
имеет размерность  $N \times N$ 
\begin{gather}\label{eq:corrected}
\tilde{F}^G_{i,j} = N_{seg} \sum_{l=1}^{N_{seg}}F^G_{i,l,j,1}, \\
\tilde{Q}_{\tau,j}= A_{j,1}^R \epsilon_{j,1} n^2 \sigma
\gamma(T_{j,1}) T_{j,1}^4,
\end{gather}
индекс 1 относится к первому сегменту тороидального элемента.

Метод был использован для моделирования теплообмена в эпитаксиальном реакторе Centura, в котором нагрев подложек обеспечивается галогенными лампами с позолоченными отражателями. Расчет проводился в рамках трехполосной модели излучения для учета спектральной зависимости оптических свойств кварца. Показано, в частности, что нагрев самих подложек производится не излучением ламп, а вторичным излучением графитового подложкодержателя. Расчет методом дискретных ординат без учета зеркального отражения оказался неудовлетворительным \cite{koch,60}.

\subsection{Фазовая граница}
\subsubsection{Рост из расплава}
При фазовом переходе первого рода расплав-кристалл выделяется скрытая теплота кристаллизации. Границу между фазами (обычно несколько молекулярных слоев \cite{interfa}) 
можно рассматривать как резкую или как тонкую переходную область с непрерывным изменением термодинамических параметров --- подходы Гиббса и Ван дер Ваальса соответственно \cite{68}. Первый более распространен, особенно в задачах с простой формой интерфейса. 
 
М. Бенеш \cite{68} перечислил материальные соотношения, относящиеся к кристаллизации: 1) свойства материала в жидкой (l) и твердой  (s) фазах: 
удельная (на единицу объема) энтальпия  $H_{\mathrm{s}} (T)$, $H_{\mathrm{l} } (T)$); удельная энтропия $S_{\mathrm{s}} (T)$, $S_{\mathrm{l}} (T)$;
удельная свободная энергия $F_{\mathrm{s}} (T) = H_{\mathrm{s}} (T) - T S_{\mathrm{s}} (T)$, \quad $F_{\mathrm{l}} (T) = H_{\mathrm{l}} (T) - T S_{\mathrm{l}} (T)$ и 2) свойства фронта кристаллизации --- интерфейса:
удельная (на единицу площади) энергия интерфейса $e (T)$; удельная энтропия $s (T)$; удельная свободная энергия интерфейса $f (T) = e (T) - T s (T)$.

Скрытая удельная теплота кристаллизации определяется как 
$L = H_{\mathrm{l}} (T^{\star}) -  H_{\mathrm{s}} (T^{\star})  $, где температура перехода $T^{\star}$ --- температура, при которой свободные энергии фаз равны  $ F_{\mathrm{l}} (T^{\star}) = F_{\mathrm{s}} (T^{\star})$.

Моделирование роста из расплава включает нахождение неизвестной границы - в наиболее сложном случае метода зонной плавки из сопряженной электромагнитно-гидродинамической задачи
 \cite{24}.

Затвердевание "`простого"' вещества определяется градиентом температуры --- это классическая задача Стефана:
область 
$\Omega$ разделена неизвестной границей $\Gamma$ на две подобласти  - жидкую  $\Omega_{\mathrm{l}}$ и твердую  $\Omega_{\mathrm{s}}$ так, что $\Omega = \Omega_{\mathrm{l}} \cup \Omega_{\mathrm{s}}  \cup \Gamma$ 
\cite{68}. Уравнение теплопроводности решается в обеих областях.
\begin{align}
\frac{\partial H_{\mathrm{l}}}{\partial t} &= - \nabla \vec{q}_{\mathrm{l}} \qquad \text{в} \qquad \Omega_{\mathrm{l}} (t), \\
\vec{q}_{\mathrm{l}} &= - \lambda_{\mathrm{l}} (T) \nabla T,\\
\frac{\partial H_{\mathrm{s}}}{\partial t} &= - \nabla \vec{q}_{\mathrm{s}} \qquad \text{в} \qquad \Omega_{\mathrm{s}} (t), \\
\vec{q}_{\mathrm{s}} &= - \lambda_{\mathrm{s}} (T) \nabla T.
\end{align}

На интерфейсе $\Gamma$ должны выполняться следующие условия:
\begin{align}
T = &\frac
{H_{\mathrm{l}} |_{\mathrm{l}} - H_{\mathrm{s}} |_{\mathrm{s}} - k_{\Gamma} e}
{S_{\mathrm{l}} |_{\mathrm{l}} - S_{\mathrm{s}} |_{\mathrm{s}} - k_{\Gamma} s},\\
(\vec{q}_{\mathrm{l}} |_{\mathrm{l}} - \vec{q}_{\mathrm{s}} |_{\mathrm{s}}) n_{\Gamma} &= v_{\Gamma} (H_{\mathrm{l}} |_{\mathrm{l}} - H_{\mathrm{s}} |_{\mathrm{s}}) - k_{\Gamma} e v_{\Gamma} - D_t e,
\end{align}
\noindent где $\vec{n}_{\Gamma}$ --- единичная внешняя нормаль  по отношению к  $\Omega_{\mathrm{s}}$, $v_{\Gamma}$  --- нормальная скорость интерфейса, $D_t$ --- производная по времени от e на $\Gamma (t)$, $k_{\Gamma} = \nabla \cdot \vec{n}_{\Gamma}$  --- средняя кривизна   $\Gamma (t)$.

Если энтальпия определяется как 
$$
H_{\mathrm{s}} = \int \limits_0^T \varrho_{\mathrm{s}} (u) c_{\mathrm{s}} (u) du, \qquad
H_{\mathrm{l}} = \int \limits_0^T \varrho_{\mathrm{l}} (u) c_{\mathrm{l}} (u) du + L,
$$
\noindent где $\varrho (T)$ и $c (T)$ - плотность и теплоемкость соответствующих фаз, разница удельных энтропий как  $\Delta s = S_{\mathrm{l}} |_{\mathrm{l}} - S_{\mathrm{s}} |_{\mathrm{s}}$, формулировка задачи Стефана упрощается \cite{68}
\begin{align}
&\varrho  (T) c  (T) \frac{\partial T}{\partial t} = \nabla \cdot (\lambda (T) \nabla T) \qquad \text{in} \qquad \Omega_{\mathrm{l}}\quad \text{and} \quad \Omega_{\mathrm{s}},\\
&\lambda_{\mathrm{s}}  (T) \frac{\partial T}{\partial n_{\Gamma}} |_{\mathrm{s}} -  
\lambda_{\mathrm{l}}  (T) \frac{\partial T}{\partial n_{\Gamma}} |_{\mathrm{l}} = L v_{\Gamma},\\
&T - T^{\star} = - \frac{\sigma (T)}{\Delta s} k_{\Gamma} - \alpha \frac{\sigma (T)}{\Delta s} v_{\Gamma}.
\end{align}
\noindent Здесь $\sigma$ --- поверхностное напряжение между фазами и  $\alpha$ --- кинетический коэффициент присоединения. 

Разница между температурами интерфейса и перехода называют кинетическим переохлаждением (эффект Гиббса-Томсона).
 
Иногда и другие задачи с неизвестной границей называют задачей Стефана.
Такие задачи впервые рассматривали Г. Ламэ и Б. П. Е. Клапейрон (1831), однако используется имя Стефана, который изучал таяние полярного льда столетие спустя 
\cite{solid}.
О задаче Стефана, усложненной конвекцией, см. 
\cite{conv} для случая ньютоновской жидкости и  \cite{nedoma} для случаев более общей реологии. 

Методы решения задачи Стефана делятся на однородные (методы сквозного счета) и методы выделения межфазной границы. Классификация очевидна и аналогична методам расчета невязких течений газа с ударными волнами. В первом подходе граница восстанавливается (реконструируется), во втором она отслеживается явно и решение ищется в двух (в трех при включении в рассмотрение  равновесной двухфазной зоны) подобластях.

Методы реконструкции основаны на отслеживании границы или  объемной фазы. В первом случае 
положение границы идентифицируется по набору маркерных точек, движение которых определяется уравнениями в локальной криволинейной ортогональной к интерфейсу системе координат; как правило,  требуется сглаживание границы.
Сложности возникают в окрестности угловых точек; подход чувствителен к вычислительному шуму.

Наиболее известные методы второго подхода - разработанный полвека назад Ф. Харлоу с коллегами метод маркеров и ячеек МАС (marker and cell) 
и метод "`объема жидкости"' VOF (volume of fluid), основанный на отслеживании доли каждой фазы в ячейке.

В классической задаче Стефана термодинамические свойства считаются постоянными, а плотности жидкой и твердой фаз --- равными. Различие плотностей, однако, может быть существенным не только для воды, но и, например для сапфира
\cite{sapph}. 
Этот случай рассмотрели М. Грибель и др.
\cite{griebel}. Сохранение массы требует равенства потоков в обеих фазах на интерфейсе 
\begin{equation}
\vec{v} \cdot \vec{n} = \left(1 - \frac{\varrho_{\mathrm{s}}}{\varrho_{\mathrm{l}}} \right) \vec{V}_{\varGamma} \cdot \vec{n} \; ,  
\end{equation}
\noindent где $\vec{v}$ --- скорость жидкой фазы и $\vec{V}_{\varGamma}$ --- скорость интерфейса. 

Температура на интерфейсе непрерывна  $T_{\mathrm{s}} = T_{\mathrm{l}}$;
для формулировки второго условия авторы рассмотрели баланс энергии как
\begin{equation}
\left[\varrho e (\vec{v} - \vec{V}_{\varGamma}) + \vec{q} \cdot \vec{n} - \vec{n}^{\mathrm{T}} \mathcal{P} \vec{n} \right]_{\mathrm{s}}^{\mathrm{l}} - =
- \gamma K \vec{V}_{\varGamma} \cdot \vec{n} \; ,
\end{equation} 
\noindent где $e = u + \vec{v} \cdot \vec{v} /2$, внутренняя  энергия жидкой и твердой фаз  $u_{\mathrm{l}} = c_{\mathrm{l}} (T - T_m)$ и $u_{\mathrm{s}} = c_{\mathrm{s}} (T - T_m) - L$  соответственно, а член в правой части связан с энергией интерфейса (эффект Гиббса-Томпсона), $\hat \tau$ - тензор напряжений в уравнениях Навье-Стокса 
\begin{equation}
\hat \tau^{ij} = -p \delta_{ij} + \mu \left(\frac{\partial v_i}{\partial x_j} + \frac{\partial v_j}{\partial x_i} \right) + \lambda \nabla \vec{v} \delta_{ij} .
\end{equation}
\noindent Подробности анализа можно найти в цитированной работе.

Описание интерфейса усложняется в случаях сегрегации компонентов расплава,  захвата чужеродных частиц движущимся фронтом кристаллизации  
\cite{bagd,z9,rosenb} или огранки кристалла. 
Явление сегрегации важно при выращивание тройных соединений полупроводников
III-V групп \cite{69}. Проблемы роста тройных соединений  GaInAs и  GaInSb связаны с большим разделением линий ликвидуса и солидуса на фазовой диаграмме и, как следствие, малым коэффициентом сегрегации In.  
Концентрация компонента на интерфейсе возрастает, диффузия и конвекция существенно влияют на качество и однородность кристалла 
\cite{69,70}. Различные дефекты (ячеистая структура, поликристалличность) наблюдаются при превышении концентрацией In некоторого критического значения \cite{71}. Температурная зависимость коэффициента диффузии влияет на возникновение неустойчивости  \cite{72}. Анализ фазовой границы должен учитывать ряд локальных взаимосвязанных явлений - тепломассоперенос, кинетику процессов на интерфейсе, переохлаждение, морфологическую устойчивость фронта кристаллизации.

\subparagraph{Огранка}
Огранка часто наблюдается при росте полупрозрачных кристаллов --- граница не отслеживает изотерму плавления, а совпадает с одной из кристаллографичеких плоскостей, кристалл имеет форму неправильного многогранника. Переохлаждение на интрефейсе может быть значительным (почти 7 К для кремния 
\cite{73}, до 20 K для германата висмута \cite{74}). 
Часто утверждается, что огранка снижает качество кристалла и ее следует избегать, поддерживая температурный градиент выше критического значения     \cite{75}. Недавно, однако, почти совершенные кристаллы германата были получены с полностью ограненным фронтом кристаллизации 
\cite{76}. Обсуждение моделирования роста кристалла с частично ограненным интерфейсом см.  в работе Вирозуба и Брендона \cite{77}.

\subsubsection{Рост из паровой фазы}

  Методы роста кристаллов  из газовой (паровой) фазы принято подразделять на химическое осаждение и физический транспорт. Первый метод используется для выращивания ЭС (двойные, тройные, четверные соединения) для приборов микро- и  оптоэлектроники  
\cite{epi,epi1}. Однако, галидная (хлоридная) эпитакия из газовой фазы может использоваться и для получения "`квазиобъемных"'  подложек широкозонных полупроводников.  

Иногда рост из паровой фазы удобно рассматривать как двухстадийный процесс, явно выделяя перенос из объема в адсорбционный слой и встраивание материала в решетку \cite{sapph}. 

\paragraph{Граничные условия}   
Ниже приведены граничные условия для уравнений массопереноса в  следующих предположениях: 1) свойства адсорбционного слоя идентичны свойствам твердой фазы; 2) рост лимитируется массопереносом из объема к поверхности: 

\noindent отсутствие полного потока инертного газа
\begin{equation}
 \rho uc_0 + J_0 = 0,
\end{equation}  

\noindent связь полных потоков компонентов и скоростей гетерогенных химических реакций 
\begin{equation} 
\rho uc_i + J_i = M_i\sum_{r=1}^{N_r}\nu_{ir}\mbox{\.w}_r\,,  i=1,\ldots,N^r_k, 
\label{eq:Fluxes} 
\end{equation}   
\noindent закон действующих масс (десорбция с поверхности происходит в условиях термодинамического равновесия)
\begin{equation} 
\prod_{i=1}^{N^r_k}X_i^{\nu_{ir}} = K_r\,,         
r=1,\ldots,N_r  \label{eq:MassAct} 
\end{equation}   
\noindent условие нормировки
\[
     \sum_{i=0}^{N^r_k}X_i = 1,
\]
\noindent где $N^r_k$ --- число газовых компонент, участвующих в реакции, индекс  "0" \rm{} обозначает инертный ("`несущий"') газ, 
$N_E$ --- число элементов,  $N_S$ --- число твердых фаз,
$N_r=N^r_k-N_E+N_S$ --- число реакций, $\nu_{ir}$ стехиометрический коэффициент $i$-го компонента в $r$-ой реакции, $\mbox{\.w}_r$ скорость  $r$-ой гетерогенной реакции, $u$ --- Стефановская скорость, $c_i$ и $M_i$ --- массовая концентрация и молярная масс $i$-го компонента,
$J_i$ --- нормальный диффузионный поток $i$-го компонента, $K_r$ --- константа равновесия $r$-ой гетерогенной химической реакции.   
В отсутствии экспериментальных данных $K_r$ оцениваются  по 
потенциалам Гиббса индивидуальных веществ   при нормальных условиях $G_i$ 
\begin{equation}
\ln K_r=\frac{1}{RT}\sum_{i=1}^{\rm N^r_k+N_s}\nu_{ir}G_i - 
      \ln\left(\frac{p_s}{p}\right)\sum_{i=1}^{\rm N^r_k}\nu_{ir}\,.
\end{equation} 

\paragraph{Сублимационный рост}

Сублимационный рост широкозонных полупроводников (SiC \cite{sic_e}, AlN) проводится, в отличие от эпитаксиального роста, в плотно закрытом контейнере.
Газовая фаза может содержать компоненты, отличные от материала кристалла, (например, при росте SiC --- Si, Si$_2$C, SiC$_2$ \cite{14,137}). Ниже рассмотрен  сублимационный  рост нитрида алюминия.

Простейший для анализа случай сублимационного роста --- конгруэнтный (бездиффузионный) перенос "`Стефановским"' ветром
\cite{s6}) достигается в опыте понижением давления \cite{wolfson}. 
Скорость роста в "`вакуумных"' условиях (ростовая ячейка помещена в специальный контейнер, в котором поддерживается давление около 10$^{-4}$ Torr) соответствует скорости роста в атмосфере азота при более высоких (на 350--400 K) температурах  \cite{s6}. В другом предельном случае преобладания инертного газа скорость роста пропорциональна разнице парциальных давлений на источнике и на затравке.

Сублимационный рост AlN можно представить как  
\[
(AlN)_{\mathrm{solid}} \xrightarrow[T_{growth} + \Delta T]{Sublimation} Al_{vapour} + \frac{1}{2}N_2 \xrightarrow[T_{growth}]{Deposition} (AlN)_{\mathrm{solid}}
\]
Метод, развитый в 1976 Г. А. Слэком и Т. Ф. МакНелли \cite{m1} (наибольший кристалл выращенный ими был длиной 10 мм и диаметром 3 мм) в настоящее время обеспечивает  рост монокристаллов диаметром до 2$^{\prime \prime}$ со скоростью до 1 мм/час \cite{m2} и высокое качество 
(ППД менее  1000 cm$^{_2}$ \cite{m3}). 

Используется как спонтанное зарождение без попытки контроля ориентации кристалла, так и затравки, обеспечивающие гомоэпитакисальный 
\cite{hart} или гетероэпитакисальный рост \cite{miyan,lu06}).
Применяются подложки из AlN, SiC  \cite{lu08,mokh03} или других материалов (сапфир, карбид тантала TaC или ниобия NbC \cite{lee07}). 
Разложение  SiC при высоких температурах может спровоцировать рост поликристаллического  AlN \cite{nov04}.
В отличие от роста монокристаллов SiC, при росте AlN не наблюдаются разные политипы: структура вюрцита 2H имеет наименьшую энергию образования \cite{bond07}.

Продолжительность процесса ограничена деградацией тигля 
\cite{wang,rojo} и изоляции из графита \cite{cai}; длительное функционирование источника требует минимизации температурных градиентов в его пределах  \cite{14,bogd}.
Размер зародыша  AlN \cite{yazdi} и морфология кристалла \cite{sitar04} зависят от температуры роста. Дислокации формируются как в процессе роста, так и на стадии охлаждения при релаксации термомеханических напряжений  \cite{def,79,oxford}. 
Влияние разориентации подложки 6H-SiC на моду роста и дефекты изучалось в работах \cite{edgar}, \cite{yak}; различные моды роста связаны с низкой подвижностью адатомов на поверхности кристалла.

Впервые одномерная модель (в пренебрежении переносом) роста AlN  предложена  в работе  \cite{s1} в предположении, что рост лимитируется разложением азота. В большинстве последующих исследований считается, что рост определяется переносом газовых компонентов \cite{s4,s5}, а скорость роста находится из уравнения Герца-Кнудсена для потока  Al на поверхности AlN; явное выражение для скорости роста получено для случая малой концентрации   Al в газовой фазе. В этих работах, однако, не определены пределы применимости приближений для режимов роста (лимитированных кинетикой или переносом), не учтен массообмен через щели и/или проницаемую стенку контейнера с окружающей средой, не рассматривалась эволюция кристалла и источника и изменение условий роста, приводящее к изменению формы фронта кристаллизации и условий формирования дислокаций и других дефектов. 

Квазистационарная одномерная модель 
\cite{s6,karp01,s8}
разработа в предположении, что 1) только  Al и N$_2$ присутствуют в газовой фазе (летучими примесями можно пренебречь); 2) скорость роста кристалла AlN определяется локальным составом газовой фазы и температурой, но не зависит от ориентации поверхности (изотропный рост); 3) испарение происходит только на поверхности источника (т.  е. вместо порошкового источника рассмотрен поликристалл). 

Модель основана на упрощенном квазитермодинамическом описании кинетики на поверхности AlN, которое применялось и к другим процессам роста  (см. \cite{s9} и приведенные там ссылки). 
Используются расширенные соотношения Герца-Кнудсена для двух газовых компонентов  Al и N$_2$
$
J_i = \alpha_i (T) \beta_i (T) (P_i^w - P_i^e).
$
\noindent Здесь, $J_i$ --- молярные потоки, $\alpha_i (T)$  --- зависящие от температуры коэффициенты прилипания, 
$\beta_i (T) = (2 \pi \mu_i R T)^{-1/2}$
\noindent --- столкновительные факторы Герца-Кнудсена, $\mu_i$ --- молярные массы, $R$ --- газовая постоянная, $ P_i^w $ --- парциальные давления на интерфейсе, $ P_i^e$ --- квазиравновессные (термодинамические) давления компонентов. 
Коэффициент прилипания Al, в силу его высокой реактивности, принят равным 1,  в отличие от  малого коэффициента прилипания азота. В работе  \cite{karp01} последний был аппроксимирован как функция температуры на основании данных по испарению AlN в вакууме  \cite{s10} (более поздние данные \cite{s11} подтвердили справедливость аппроксимации)
$\alpha_{N_2} (T) = (3.5 \exp{-30000/T})/(1 + 8 \cdot10^{15} \exp{55000/T})$.

Давления $ P_i^e$  подчиняются закону действующих масс
$
\left(P_{Al}^e\right)^2 P_{N_2}^e = K (T),
$
\noindent где $K (T)$ --- константа равновесия реакции
$2Al + N_2 \Leftrightarrow 2AlN(s).$
Модель была верифицирована сравнением с экспериментальными данными по скорости роста AlN в зависимости от температуры и давления  \cite{s8} и используется в программном обеспечении Virtual Reactor$^{TM}$ \cite{vr1,143}.  
Моделирование показало, в частности, высокую чувствительность роста к распределению температуры в контейнере и важность адекватного описания теплообмена.

Если бы рост происходил в герметичном химически инертном контейнере, установившееся давление обеспечивало бы постоянство разницы в количестве атомов алюминия и азота --- массообмен твердое тело --- пар происходит стехиометрически на поверхности  AlN.
Из-за существования щелей и проницаемости графитовых стенок окружающая среда влияет на рост.
В равновесии $J_i = 0, P_i^e = P_i^w$,  давления компонентов находятся из системы
\begin{eqnarray}
\left(P_{Al}^w\right)^2 P_{N_2}^w &= K (T)\\
P_{Al}^w + P_{N_2}^w &= P,
\end{eqnarray}
\noindent где  $P$ --- полное давление в контейнере. Анализ показывает, что если 
$P > P^{\star} (T) = 3/2 [2K (T)]^{1/3}$,  где $P^{\star}$ обозначает критическое давление, существуют два решения для смесей, обогащенных  атомами  Al или  N.
Поскольку происходит массообмен с окружающей средой (азот), следует выбрать второе решение.
Если  $P = P^{\star} (T)$,
\noindent существует единственное решение, соответствующее стехиометрическому составу,
а если  $P < P^{\star} (T)$ --- решения нет.

Это означает, что в герметически закрытом контейнере равновесное давление всегда выше критического.
В негерметичном контейнере равновесие невозможно, если давление ниже критического: в этом случае испаряются источник и затравка, а смесь 
Al/N$_2$ выходит из контейнера.
Массообмен с окружающей средой происходит через щель между стенкой контейнера и крышкой, давления внутри и снаружи близки при высоком внешнем давлении, но значительно отличаются при уменьшении последнего, завися от гидравлического сопротивления зазора $\zeta$ (скорость роста при этом падает).
При достаточно низком внешнем давлении 
$P_a^0$ начинается испарение кристалла.
Значение  $P_a^0$ уменьшается с гидравлическим сопротивлением 
(при достаточно высоком значении $\zeta$ испарения кристалла не происходит при сколь угодно малом внешнем давлении).
Скорость роста/испарения определяется локальным пересыщением.
Поток на интерфейсе можно оценить как \cite{z10aln,z12}
\begin{equation}
J \approx \frac{\left(P_{Al}^w\right)^2 P_{N_2}^w /K(T) - 1}{4/[3 \beta_{Al} (T) P_{Al}^w]  + 1/[3 \alpha_{N_2} \beta_{N_2} (T) P_{N_2}^w]}.
\end{equation}
\noindent Здесь в числителе локальное пересыщение, а знаменатель соответствует  локальному кинетическому сопротивлению.
На периферии пересыщение оказывается отрицательным из-за утечки через кольцевую щель, приводя к грибообразной форме кристалла \cite{z11,z12}.

\subsection{Электромагнитные явления} 
  Степень связи электромагнитных явлений с другими процессами различна в разных методах роста. СВЧ нагрев можно рассматривать независимо от теплообмена.
Возможность независмого рассмотрения магнитного поля, используемого для подавления турбулентных пульсаций или для интенсификации перемешивания расплава, определяется магнитным числом Рейнольдса $Re_m$ 	\cite{mf1}. 
В методе зонной плавки электромагнитная задача связана не только  с расчетом течения расплава, но и с переносом примеси вблизи фронта кристаллизации через электрическую проводимость  \cite{24,nacke}.
  
	Обычно диэлектрическую проницаемость $\epsilon$ и магнитную проницаемость $\tilde{\mu}$ принимают постоянными (линейная изотропная среда), поэтому уравнения Максвеела записываются в виде 
\begin{equation}
\nabla \cdot (\epsilon {\bf E}) = \rho_e,
\nabla \cdot {\bf B} = 0,
\nabla \times {\bf E} = - \frac{\partial {\bf B}}{\partial t},
        \nabla \times \left(\frac {{\bf B}}{\tilde{\mu}} \right) = {\bf j} + \epsilon \frac{\partial {\bf E}}{\partial t}.
\end{equation}
    
   В силу соленоидальности ${\bf B}$ можно ввести векторный потенциал ${\bf A}$    
   $\nabla \times {\bf A} = {\bf B}$, 
   тогда закон Фарадея запишется как 
\begin{equation}
\nabla \times \left({\bf E} + \frac{\partial {\bf A}}{\partial t}  \right) = 0,
\end{equation}
   что позволяет ввести скалярный потенциал $\phi$ 
\begin{equation}
{\bf E} = - \nabla \phi - \left( \frac{\partial {\bf A}}{\partial t} \right).
\end{equation}

    Векторный потенциал подчиняется закону Ампера 
    $$\nabla \times \nabla \times  {\bf A} + \epsilon \tilde{\mu} \frac{\partial}{\partial t}\left( 
    \nabla \phi + \frac{\partial {\bf A}}{\partial t} \right) = \tilde{\mu} {\bf j},$$ 
        \noindent где плотность тока  ${\bf j}$  - сумма компонеты, индуцированной внешним генератором  ${\bf j}_{ext}$ и вихревой компоненты, индуцированной в нагрузке  ${\bf j}_{eddy}$.  Это уравнение преобразуется с помощью калибровки Лоренца
$\nabla \cdot {\bf A} +\epsilon \tilde{\mu}  \partial \phi /\partial t = 0$ к
\begin{equation}
\nabla ^2 {\bf A} - \epsilon \tilde{\mu} \frac{\partial ^2 {\bf  A}}{\partial t^2} = - \tilde{\mu} {\bf j}.
\end{equation}
Подобное уравнение получается для скалярного потенциала 
\begin{equation}
\nabla ^2 \phi - \epsilon \tilde{\mu} \frac{\partial ^2 \phi}{\partial t^2} = - \frac{\rho_e}{\epsilon}.
\end{equation}
   
\paragraph {СВЧ нагрев}
СВЧ нагрев \cite{wu05} при сублимационном росте, в методе Бриджмена и ряде других слабо связан с задачей теплообмена --- только через температурную зависимость материальных свойств. Отсутствие свободных зарядов  $\rho_e = 0$ позволяет избавиться от скалярного потенциала.
Электромагнитное поле обычно считается гармоническим с амплитудой и фазовым сдвигом, зависящими от положения  $\Phi (t) = \Phi_m e^{i(\omega t +\phi)}$ \cite{wias}.  
Тогда  ${\bf j}_{eddy} = \sigma {\bf E }_{eddy} = -i \sigma  \omega_{ext} {\bf A}$. 

Записывая ${\bf A} = {\bf A}_0 e^{i  \omega_{ext} t}$  имеем:
\begin{equation}
\nabla ^2 {\bf A}_0 - \epsilon  \omega_{ext} ^2 \tilde{\mu} {\bf A}_0 - i  \omega_{ext} \sigma \tilde{\mu} {\bf A}_0 = \tilde{\mu} {\bf j_0},
\end{equation}  
\noindent где  $\sigma$ --- электропроводность, ${\bf j}_0$ --- амплитуда тока, приложенного к системе; вторым членом можно пренебречь, т. к. в ростовых установках $L << c/ \omega_{ext}.$
Мгновенное выделение энергии
$$- {\bf j} {\bf E} = \sigma \left( \frac{{\bf A}}{\partial t} \right)^2. $$
Поскольку время релаксации значительно меньше характерного теплового, тепловыделение дается усредненной по времени величиной 
\begin{equation}
Q = \frac{1}{2} \sigma  \omega_{ext}^2 |A_0|^2.
\end{equation}

Нагревательные системы часто (приблизительно) осесимметричны:  пренебрегая спиральностью индукционной катушки, запишем  ${\bf j}_0 = j_0 {\bf e}_{\theta}.$
Тогда требуется найти только азимутальную компоненту  $A_{\theta}$ 
векторного потенциала из уравнения
$$\frac{1}{r} \frac{\partial}{\partial r} \left( r \frac{\partial A_{\theta}}{\partial r} \right) - \frac{A_{\theta}}{r^2} + \frac{\partial^2 A_{\theta}}{z^2} - i  \omega_{ext}  \sigma \tilde{\mu} A_{\theta} = - \tilde{\mu} J_{\theta}. $$ 

 \paragraph{Действие магнитного поля на течение расплава} 
Использование магнитного поля для подавления турбулентных пульсаций при выращивании Si методом Чохральского является обычным (впервые, однако, оно применено  при росте  InSb \cite{mf1}). Движение расплава поперек магнитных линий порождает индукционный ток ${\bf J } = \sigma ({\bf V \times B})$.   В итоге появляется сила Лоренца ${\bf F = J \times B}$, действующая на расплав. 

Закон Ома ${\bf j} = \sigma ({\bf E} + {\bf V} \times {\bf B})$ позволяет исключить электричекий ток из закона Ампера, что (в предположении $\sigma = const$) после применения операции вихря дает 
\begin{equation}
\nabla \times  \left(\nabla \times \frac{\bf B}{\tilde{\mu}} \right) = \sigma \nabla \times ({\bf E} + {\bf V} \times {\bf B}).
\end{equation} 

\noindent Использование уравнения Фарадея для исключения ${\bf E}$  дает уравнение переноса магнитного поля \cite{huang},  которое показывает связь магнитного поля {\bf B} и поля скорости {\bf V}. 
\begin{equation}
\frac{\partial {\bf B}}{\partial t} + ({\bf V} \cdot \nabla) {\bf B} = \frac{1}{\tilde{\mu} \sigma} \nabla^2 {\bf B} + {\bf B} \cdot (\nabla {\bf V}).
\end{equation}

Член с Лапласианом соответствует диффузии магнитного поля, а последний член - растяжению магнитных силовых линий из-за градиента скорости.
Мерой связи поля скорости ${\bf V}$ и  магнитного поля  ${\bf B}$ 
служит магнитное число Рейнольдса  
$Re_m = \tilde{\mu} \sigma V L = \frac {V L}{\eta}$, где $\eta$ - коэффициент магнитной диффузии. При малых $Re_m$ расчет магнитного поля отщепляется от гидродинамической задачи: магнитное поле не зависит от скорости (но влияет на течение)  
\begin{equation}
\frac{\partial {\bf B}}{\partial t} = \frac{1}{\tilde{\mu} \sigma} \nabla^2 {\bf B}.
\end{equation}
 В случае приложения статического магнитного поля к течению с малым $Re_m$ имеем $\nabla^2 {\bf B} = 0$.
Сила Лоренца записывается как
$Ha^2  \nabla \times ({\bf B}) \times  {\bf B} $, $Ha = \left( \sqrt{\sigma \tilde{\mu}} B_0 L_0\right)$ --- число Гартмана.

На практике используется магнитное поле разной природы --- стационарное (горизонтальное, вертикальное, неоднородное) или нестационарное (вращающееся, бегущее, пульсирующее).
Заметим, что для генерации одной и той же плотности электромагнитной силы 
 F$_{EM}$ требуется значительно большая индукция  B стационарного поля ($B \approx  \sqrt{F_{EM} /(\sigma u)}$), чем в нестационарном случае  ($B \approx \sqrt{F_{EM}/(\sigma B^2 \pi f L)}$).
  
Высокочастотное магнитное поле проникает на глубину, определяемую скин-эффектом - вводится безразмерный {\em параметр экранирования}  
 $R_{\omega} = \tilde{\mu} \sigma  \omega_{ext} L^2,$ 
 позволяющий записать
\begin{equation}
R_{\omega} \frac{\partial {\bf B}}{\partial t} = \nabla^2 {\bf B}.
\end{equation}
  \noindent Параметр экранирования можно интерретировать как  $$R_{\omega} = \frac{\text{диффузионное время}\quad {\bf B} }{\text{период колебаний}\quad {\bf B}_{ext)} }. $$
 Введение глубины проникания магнитного поля   $\delta_m^{\star} = \sqrt{2/\tilde{\mu} \sigma \omega_{ext}}$ позволяет записать $R_{\omega} = 2 (L/\delta_m^{\star})^2$
  \cite{huang}.

Частота внешнего магнитного поля для перемешивания расплава выбирается на основе времени реакции течения расплава.  В случае вращающегося магнитного поля 
$B = B_0(\cos (\theta - \omega_{ext} t) {\bf e}_r - \sin (\theta - \omega_{ext} t) {\bf e}_{\theta}) $ 
отношение электромагнитных сил к вязким характеризуется магнитным числом Тэйлора 
 $Ta_m = \frac {\sigma \omega_{ext} B_0^2 L_0^4}{2 \rho \nu^2}$.  

Для магнитного поля низкой частоты эффект экранирования слаб 
 $R_{\omega} \ll  1$: глубина проникания значительно больше размера области расплава.  При росте кристаллов обычно $R_{\omega} \gg R_m$ и низкочастотное приближение справедливо ($\nu/L_0$  --- масштаб скорости), если частота удовлетворяет условию \cite{huang}
\begin{equation}
\frac{\nu}{L_0^2} \ll \omega_{ext} \ll \frac{1}{\tilde{\mu} \sigma L_0^2}.
\end{equation} 

Используются и другие упрощения (например, идеально проводящие или изолирующие стенки \cite{griesse}), требующие  аккуратности --- так, нельзя считать границу расплав-кристалл изолирующей или пренебрегать слабым электрическим током в кристалле \cite{ivan}. 

Альтернативная формулировка уравнений магнитной гидродинамики 
 (формулировка {\em скорость - ток}) \cite{griesse,meir} использует плотность тока вместо индукции магнитного поля как неизвестную переменную. Магнитное поле индуцируется токами ${\bf j}$ во внешних проводниках и в расплаве согласно закону Био-Савара
\begin{equation}
{\bf B ({\bf x})} = -\frac{\tilde{\mu}}{4 \pi} \int_{\mathbb{R}^3} \frac{\bf x - y}{|{\bf x - y}|^3} \times {\bf j ({\bf y})} d^3 {\bf y}.
\end{equation}
Это соотношение обеспечивает  $\nabla \times  
{\bf B} = \tilde{\mu} {\bf j}, \nabla \cdot {\bf B} = 0$ в ${\mathbb{R}^3}$.

\subsection{Оценка качества выращенного кристалла} 
Моделирование тепломассообмена дает скорость роста (и, т. о., форму кристалла) и состав. Для оценки качества кристалла необходимо  исследовать  формирование дефектов --- нарушений трансляционной симметрии кристалла  - и их эволюцию как в процессе роста, так и на стадии охлаждения.  Одна из возможных схем классификации дефектов основана на их размерности \cite{duffar}: нульмерные (точечные) дефекты  (вакансии, собственные межузельные атомы, примесные атомы замещения  или  внедрения  \cite{ammon,valek}), линейные (дислокации и их коллективные структуры, например, ячеистые  \cite{rud07}), двумерные (границы зерен, двойники, малоугловые границы для бинарных соединений полупроводников A$_3$B$_5$ --- GaAs, InP \cite{mullin}), локализованные трехмерные (преципитаты, скопления вакансий, кластеры Si для кремния, микропузырьки для оптических кристаллов, микротрубки и включения иных политипов для карбида кремния, трещины для тройных соединений), диффузные трехмерные (химическая неоднородность, остаточные напряжения).  Степень связи оценки качества кристалла с расчетом тепломассообмена варьируется (Табл. 5).
Например, неоднородность состава и концентрация микропузырьков в оптических кристаллах определяется одновременно со скоростью роста. Вычисление термоупругих и концентрационных напряжений в случае, когда нет пластической деформации, является стационарной задачей в силу большой скорости звука в твердых телах и может проводиться независимо, так же как и нестационарный анализ эволюции преципитатов кислорода и скоплений вакансий в кремнии.

\paragraph{Примеси}

Не все примеси вредны. Иногда утвержается, что хороший кристалл кремния должен содержать минимальное количество кислорода и углерода 
 \cite{jin} или что "примеси встраиваются в кристалл, снижая его качество" {\ }  \cite{wias}. На деле содержание кислорода в кремнии должно быть в пределах  $(1-4) \cdot 10^{17}$ atoms/cm$^3$ \cite{vegad} 
поскольку кислород 1) увеличивает жесткость кристаллической решетки, что стабилизирует кремниевые подложки, и 2) служит центрами  геттерирования для нежелательных примесей.

Взаимодействие расплава со стенками тигля приводит к поступлению кислорода в расплав согласно реакции 
$(SiO_2)^{solid} \longrightarrow (Si)^{liquid} + 2 (O)^{gas}$   \cite{karlb}. Растворимость кислорода в кремнии зависит от температуры \cite{hir}  и от концентрации бора  \cite{abe}.  Кислород переносится расплавом и частично испаряется со свободной поверхности в виде окиси кремния, частично встраивается в растущий кристалл.  Коэффициент сегрегации  полагается равным 1 - граничное условие на интерфейсе  формулируется как нулевой нормальный градиент 
концентрации.  Реакция газообразной окиси кремния с графитовыми элементами установки дает карбид кремния и окись углерода, которые частично удаляются с потоком аргона \cite{lan}.
Перенос кислорода в расплаве и окиси кремния в потоке инертного газа  рассчитывается с использованием полей скорости, температуры и турбулентной вязкости, полученных при расчете тепломассообмена.

Уровень примесей критичен в случае объемных монокристаллов нитридов III группы, используемых для производства подложек для полупроводниковых приборов. Наилучшие подложки для приборов на основе 
GaN - GaN и  AlN. 
Последние предпочтительнее для приборов с высоким содержакнием  Al, например, 
УФ светодиодов (AlN прозрачен для длин волн до  200 нм), полевых транзисторов, вертикальных и горизонтальных диодов Шоттки и  p-i-n диодов \cite{luo02}
(непреднамеренно легированные кристаллы AlN  --- изоляторы\footnote{Концентрация кислорода в кристаллах AlN, выращенных сублимационным методом, высока \cite{herro}; однако, большая часть атомов кислорода встроена не как мелкий донор, а как примесь, образующая комплексы с точечными и протяженными дефектами, расположенные глубоко в запрещенной зоне
\cite{freitas1,freitas,schultz}. 
При концентрации кислорода ниже критического уровня формируются вакансии	  Al, при более высоких концентрациях образуется дефект на основе октоэдрически координированных атомов Al \cite{epi}.
		Кислород сильнее влияет на теплопроводность AlN из-за рассеяния фононов на  дефектах, природа которых зависит от концентрации кислорода \cite{epi,kazan1}. Так, эффективным центром рассеяния фононов является  дефект замещения вакансии алюминия \cite{rojo}.
} 
в отличие от кристаллов GaN, характеризуемых высокой концентрацией свободных электронов  
\cite{freitas}\footnote{
Так, в кристаллах  GaN, выращенных гидридной газофазной эпитаксией, непреднамеренное легирование n-типа имеет источником кремний и кислород, испаряющиеся из кварцевых элементов реактора \cite{paskova10}. Компенсационное легирование атомами железа позволяет добиться полуизолирующих кристаллов \cite{vaudo} с наименьшим уровнем концентрации свободных носителей $5 \cdot 10^{13}$ cm$^{-3}$ \cite{p13}.
}). 
\paragraph{Дефекты}
В случае чисто упругой деформации кристалла напряжение на любой стадии роста может быть вычислено независимо. Термоупругая задача является трехмерной даже для осесимметричных кристаллов, за исключением случая специальной ориентации главных осей \cite{78,79,80}. Напряжение, вызванное градиентами температуры (а также неоднородностью состава тройных соединений) может приводить к растрескиванию хрупких кристаллов \cite{81}. Композиционное напряжение в тройных кристаллах, содержащих Ga и In, может значительно превышать термоупругое, поскольку радиусы атомов  Ga и In отличаются на 12\% \cite{70}.

Превышение напряжением критического значения в ковких кристаллах приводит к ползучести. Модели эволюции дислокаций обычно оперируют зависимостью скорости пластической деформации от девиатора тензора напряжений 
и параметров дислокаций (вектор Бюргерса, скорость и т. д.)
 \cite{82,zhm_str} и уравнением для изменения ППД  \cite{75,83,84}.
Дислокации присутствуют практически во всех кристаллах, другие дефекты специфичны для определенных материалов. Например, двойники и малоугловые границы важны для полупроводниковых кристаллов $A_3B_5$ (GaAs, InP), 
включения иного политипа наблюдаются в  SiC, встраивание микропузырьков \cite{85} губительно для оксидов, галидов и  других оптических кристаллов.

Встраивание первичных точечных дефектов в кристалл кремния и их эволюция, включая формирование комплексов вакансий и преципитатов кислорода, определяются интерфейсом (форма и температурный градиент), наличием примесей, тепловой историей. Анализ этих вопросов можно найти в обзоре 
 \cite{86}.

\section{Верификация}
Верификация --- проверка правильности реализации модели и адекватность модели реальности. В англоязычной литературе используются  термины verification и validation соответственно \cite{97,derby1}.

\subsection{Верификация корректности реализации}  
Это математический вопрос  \cite{97}. Источники ошибок  перечислены Оберкампфом и Трукано \cite{100}: 
недостаточное разрешение (пространственное и временное), недостаточная сходимость итерационного процесса,  ошибки округления, ошибки программирования. 
Ошибки последнего типа особенно трудно найти, когда ПО дает 
"`умеренно неверные"' результаты  \cite{100}. 
Исследование Хаттона \cite{101} обнаружило поразительно большое число таких ошибок --- в более чем 100 регулярно используемых научных программах (исследовательских и  коммерческих).

Верификация осуществляется сравнением с эталонными решениями трех сортов: 1) аналитические решения, включая полученные так называемым {\em методом изготовленных решений} \cite{ms1,ms2,ms3}; 2) решения обыкновенных дифференциальных уравнений (ОДУ); 3)  решения уравнений в частных производных. 

Метод изготовленных решений прост, но утомителен: 
выбранное аналитическое  ``точное'' решение $\phi_{ex} (t,x,y,z)$ 
подставляется в уравнения для получения правых частей 
$f_{ex} = L(\phi_{ex}, \phi_{ex}^{\prime}, \phi_{ex}^{\prime \prime})$ и необходимых граничных и начальных условий. 
Его полезность для суждения о методах, используемых для решения реальных задач, однако, определяется близостью выбранного решения реальным явлениям. 
Наличие аналитического решения значительно уменьшает время тестирования нестационарных программ, позволяя ограничить интегрирование коротким временным интервалом.

Использование аналитических и эталонных решений ОДУ для тестирования многомерных программ использует свойства симметрии: можно решать одномерную задачу (имеющую цилиндрическую или сферическую симметрию\footnote{
Например, Г. Патерсон в 1952 получил аналитическое решение задачи Стефана в цилиндрических координатах. Линейный источник тепла $Q$ активирован в момент времени $t = 0$ на оси симметрии бесконечного тела, имеющего температуру  $T_{\mathrm{s}} < T_{\mathrm{m}}$. 
Решение Патерсона имеет вид \cite{solid}  ($\mathrm{Ei} (x)$ - экспоненциальная интегральная функция):
положение интерфейса $R (t) = 2 \lambda \sqrt{\alpha_{\mathrm{s}} t}$; температура жидкой и твердой фаз 
\[
T_l (r, t) = T_{\mathrm{s}} + \frac{T_{\mathrm{m}} - T_{\mathrm{s}}}{\mathrm{Ei} (- \lambda^2 \alpha_{\mathrm{s}}/\alpha_{\mathrm{l}})} \ \mathrm{Ei} \left( - \frac{r^2}{4 \alpha_{\mathrm{s}} t} \right),
T_s(r, t) = T_{\mathrm{m}} - \frac{Q}{4 \pi \kappa_{\mathrm{s}}} \left[ \mathrm{Ei} \left(- \frac{r^2}{4 \alpha_{\mathrm{s}} t} \right) - \mathrm{Ei} (\lambda^2)  \right].
\]

Параметр $\lambda$ находится из уравнения 
\[
-\frac{Q}{4 \pi} {e}^{- \lambda^2} + \frac{\kappa_{\mathrm{l}} (T_{\mathrm{m}} - T_{\mathrm{s}})}{\mathrm{Ei} (- \lambda^2 \alpha_{\mathrm{s}}/\alpha_{\mathrm{l}})} e^{- \lambda^2 \alpha_{\mathrm{s}}/\alpha_{\mathrm{l}}} = \lambda^2 \alpha_{\mathrm{s}} \varrho L.
\]

}), используя двумерные или трехмерные расчетные сетки.

Известен ряд многомерных эталонных решений: течение в полости с подвижной верхней крышкой 
\cite{lid,lid1,lid4}, ламинарная конвекция в квадратной \cite{102,103,nc2}  и кубической  \cite{nc3} полостях, течение над ступенью с теплообменом \cite{104}, упрощенные схемы течения расплава в методе Чохральского \cite{cz1},  осесимметричная задача о тепломассобмене при химическом осаждении кремния из силана в реакторе с вращающимся диском  \cite{43}.

\subsection{Верификация адекватности модели} 
Это второй этап оценки достоверности моделирования. Заметим, что не следует опускать первый: само по себе успешное сравнение с экспериментальными данными недостаточно в силу ограниченности числа тестов,
"графического"{\ } характера сопоставления в большинстве работ (ведущего к сомнительным утверждениям вроде
 {\em semi-quantitatively identical results}  --- "полуколичественно идентичные результаты"{\ } \cite{chen08a}) и возможной компенсации ошибок. 
 
В силу сложности  исследования процессов в реальных ростовых установках используются упрощенные конфигурации, позволяющие  изучать те или иные аспекты проблемы  \cite{100}.  Несколько таких задач приведено в работе  \cite{105}, включая т.н. "цилиндрическую конфигурацию"{\ } (цилиндрический контейнер с  коаксиальным цилиндрическим нагревателем) и "полузона" {\ }(``half zone'') \cite{106,minak}, которая имитирует жидкий мост в методе зонной плавки. 

Большинство экспериментальных  исследований направлены на характеризацию выращиваемых кристаллов 
 \cite{char} - разнообразные методы позволяют изучать поверхностные и объемные дефекты  \cite{107}.  
Такие техники как микроскопия фазового контраста, дифференциальная интерферометрическая микроскопия, атомная силовая микроскопия, сканирующая  туннельная микроскопия позволяют визуализировать растущие ступени нанометровой высоты. В специальных случаях - при кристаллизации белков - возможно даже отслеживать поведение индивидуальных "единиц роста"{\ } в силу значительного размера макромолекул и большого характерного времени \cite{108,109,110}. Формирование и пространственное распределение дислокаций, других дефектов решетки, химических неоднородностей в кристаллах исследуется с помощью рентгеновской топографии, 
томографии с использованием лазерного луча или  катодолюминесценции, ренгеновской флюоресценции и т. п.  

Серьезные технические трудности препятствуют визуализации течения и измерениям  {\em  in situ} температуры и поля концентраций. Одна из причин - экстремальные условия роста (например, рост монокристаллов GaN 
из расплава азота в галлии проводится при температуре около 1800 K и давлении до  15 000 бар \cite{g4,g5}, а сублимационный рост  AlN и SiC  - при температуре выше  2500 K). Более того, и аммонотермальный,  и сублимационный рост проводится в герметически закрытых контейнерах. 
Исследование кинетики кристаллизации в растворе также затруднительно  \cite{interf}.

Визуализация течения непрозрачного расплава возможна с помощью радиоактивных меток 
\cite{113} или рентгеновской радиографии \cite{115,115a}, основанной на зависимости поглощения от плотности материала.  Более сложным является случай сплава, т. к. вариации как температуры, так и концентрации меняют плотность и, следовательно, поглощение. 
Если коэффициент поглощения одного из компонентов значительно больше, чем остальных, можно визуализировать распределение концентрации  \cite{116}. Рентгеновскую радиографию можно использовать для отслеживания фронта кристаллизации  \cite{117}, конфокальная сканирующая лазерная микроскопия позволяет измерить переохлаждение расплава на грани кристалла  \cite{73}.
Для получения формы и положения фронта кристаллизации  используется периодическое возмущение условий роста  \cite{duffar}.

Уникальный метод непосредственного  измерения температуры на интерфейсе, основанный на прозрачности кристалла германата висмута  $Bi_4Ge_3O_{12}$ и непрозрачности расплава, позволил Голышеву и др. \cite{74,118} изучить {\em in situ} переохлаждение расплава на фазовой границе  в методе осевого теплового потока. 
 
В немногих работах приводятся надежные результаты исследования распределения температуры в ростовой установке. Авторы  \cite{119} измерили градиент температуры в растворе фосфорной кислоты при выращивании кристаллов  AlPO$_4$ и GaPO$_4$  при  500 K. Измерения флуктуаций температуры, вызванных термокапиллярной конвекцией в слое жидкого металла  (Si, Mo),  проведены как в земных условиях, так и при пониженной гравитации  (см., например,  \cite{121,122}). 

Видимо, наиболее полные и надежные исследования распределения температуры при росте кремния методом Чохральского провели Г. Мюллер с коллегами  \cite{28,123} с помощью термопар и оптических датчиков, несомненное превосходство последних позволило измерить высокочастотную область турбулентного спектра.
Авторы также разработали сенсоры для измерения концентрации кислорода как в расплаве, так и в газовой фазе 
 \cite {123}. 
Известен метод отслеживания фазовой границы на основе различия электропроводности кристалла и расплава 
 \cite{wadley}:  индуцированный ток в пределах элемента, содержащего интерфейс, зависит от доли жидкой фазы. Очевидно, его нельзя использовать для роста кристаллов диэлектриков, но метод применим к исследованию кристаллизации полупроводников: например, электропроводность арсенида галлия меняется в 15 раз при плавлении  \cite{glazov}. 

Трудности исследования реальных процессов роста кристаллов стимулируют физическое моделирование.
Так, для изучения метода Чохральского использовали силиконовое масло  \cite{124},
воду, спирт, расплав  NaCl-CaCl$_2$, сплав Вуда \cite{pol01,son}. Л. Горбунов и др.  \cite{125} применили эвтектическую смесь InGaSn с температурой плавления  285 K. Подбирая размер тигля, магнитную индукцию, скорости вращения "кристалла" {\ } и тигля, авторы добились равенства 4 критериев подобия (чисел Грасгофа, Гартмана, двух чисел Рейнольдса --- по вращению кристалла и тигля)  в физическом эксперименте и реальном росте кристаллов кремния диаметром 200 мм. Специальные устройства эмулировали потери тепла излучением свободной поверхности.
Однако, вряд ли возможно имитировать эффект Марангони и сдвиговое напряжение, вызванное потоком газа; граница "кристалла" {\ } в эксперименте с    InGaSn  была плоской, в то время как интерфейс кремния имеет значительную кривизну. 

Основной экспериментальной информацией для верификации численных методов остается форма фазовой границы.

\subsection{Оптимизация режимов роста и ростовых установок}
В некоторых примерах оптимизации режимов выращивания монокристаллов или ростовых установок число контрольных параметров мало (например, положение и  мощность нагревателя в росте из расплава 
\cite{75,134} или массовые расходы газов и скорость вращения подложкодержателя в эпитаксиальном реакторе \cite{135}) и потому можно обойтись без регуляризации задачи. Эта процедура обязательна, когда число параметров значительно   \cite{9}, как, например, для контроля состава  тройных соединений  $Al_xGa_{1-x}As$ в газофазной эпитаксии \cite{136} 
или при оптимизации формы тигля для сублимационного роста монокристаллов  SiC  \cite{14,137}. Отметим также применение сопряженного метода для оптимизации распределения граничного теплового потока в методе направленной кристаллизации и методе Бриджмена  \cite{138} и оптимизацию температуры поверхности кристалла в методе Чохральского  \cite{139}.

Поскольку решение прямой задачи трудоемко, на начальных этапах отимизации часто используют т.н. "суррогатные" {\ } модели  \cite{140}. Такие  модели получают либо из физических соображений (например, уменьшение пространственной размерности задачи), либо формальными процедурами как модели "черного ящика"  {\ } с  помощью многопараметрической аппроксимации, регрессионных методов, нейронных сетей, кригинга и т. п.  \cite{141}).
Заметим, что последний подход позволяет сочетать данные расчетов и экспериментов.

Параметрическое геометрическое моделирование, которое сводится к деформации кривых/поверхностей, существенно ограничивает пространство поиска - следует допустить топологические изменения в системе для оптимизации всей ростовой установки.  Очевидно, в настоящее время в этом отношении моделирование роста кристаллов значительно уступает более зрелым областям, например, оптимизации в аэрокосмической сфере 
 \cite{141,142}.

\section{Программное обеспечение (ПО)}
\subsection{Требования}
Моделирование роста кристаллов смещается из исследовательских институтов в промышленность в силу ряда причин:
доступность открытого и коммерческого ПО для решения задач гидродинамики и теплообмена \cite{87}; дешевизна мощных вычислительных систем; 
нежелание раскрывать конфиденциальную информацию; возможность использовать ПО в каждодневной работе.

Однако "коммерческое ПО не может эффективно использоваться в промышленности без поддержки специалистов по механике жидкости" {\ }  \cite{88}. В начале 2000 ЕС инициировал огромный проект (более 40 организаций из 11 стран)   QNET-CFD \cite{88}, целью которого было не проведение исследований, но анализ практики использования ПО в ряде областей механики жидкости и тепломассообмена (внешняя аэродинамика, горение, химическая промышленность, окружающая среда, течения в турбинах) и  выработка рекомендаций.
Моделирование роста кристаллов, основанное на вычислительной гидродинамике и вычислительном теломассообмене, 
наследует их проблемы, добавляя неизвестную фазовую границу, огранку кристаллов, анизотропию их свойств, формирование и эволюцию точечных и протяженных дефектов. Необходимо ПО, которое "экранировало"{\ } бы инженеров от вычислительных аспектов, позволяя сконцентрироваться на стоящих перед ними проблемах.

Создание таких симуляторов роста кристаллов возможно либо разработкой оболочки, обеспечивающей адаптацию коммерческого ПО общего назначения, либо проектированием специализированного ПО.
Первый подход к настоящему времени успешен только для  эпитаксии (Табл. 6). Моделирование роста монокристаллов труднее, т.к. требует решения междисциплинарной задачи в многоблочной области, а иногда и использования разных моделей в пределах одной подзадачи (например, турбулентное течение расплава  и ламинарное течение газа в методе Чохральского).
 
В Табл. 7 сравниваются требования к ПО, предназначенному для научных исследований и для решения инженерных задач. 

\paragraph{Может ли ПО быть "дружелюбным"?} 
Утверждается, что термин "дружелюбный" {\ } применительно к ПО не допускает количественной оценки 
 \cite{126}. Это не так - например, можно определить среднее время освоения ПО  или число операций/время, необходимое для задания геометрии задачи и спецификации условий роста.
Разработчики ПО в силах 1) использовать интуитивно понятный графический интерфейс; 2) минимизировать действия пользователя для формулировки задачи; 3) использовать алгоритмы, которые не требуют подстройки в процессе расчетов и 4) единицы физических величин и переменные,  принятые в эксперименте. 

Например, возможно автоматическое определение блоков из "проволочной" {\ } геометрии, заданной пользователем, 
определение внутренних (межблочных) границ, корректировка геометрии задачи из-за роста кристалла и/или движения нагревателя, в том числе топологические изменения (новые блоки и новые внутренние границы), генерация расчетной сетки.

Полная автоматизация невозможна. Наиболее трудоемкий этап -  спецификация геометрии  \cite{130}).
Импорт из программ компьютерного  дизайна (CAD) требует "ремонта"{\ } данных \cite{131}: многие форматы  CAD не содержат отношения соседства, поэтому необходимо восстановить топологию модели  \cite{132a}, что затрудняет наличие ошибочных зазоров между элементами сетки   \cite{132,132b}. 
Причиной несогласованности в данных  CAD (например, грань, ребра которой не составляют замкнутую цепь) является различие в допустимой погрешности, принятой в CAD, и требуемой для генераторов сеток  \cite{130}. В настоящее время ручная "починка" {\  } (с помощью интерактивных инструментов \cite{133}) неизбежна для сложной геометрии. 

В идеале ПО должно представлять собой черный ящик, не требующий вмешательства пользователя и обеспечивающий получение решения, которое от пользователя не зависит ( {\em user-independent} - W.G. Habashi et al., \cite{habashi}).
Цена надежности --- эффективность, поэтому необходима адаптивность различной природы: например, адаптация сетки
 \cite{127,128}, использование для решения больших разреженных систем уравнений адаптивных полиалгоритмов (упорядоченый набор алгоритмов, от быстрого ненадежного к медленному надежному) с автоматическим переключением  \cite{129}. 

\subsection{Параллельные вычисления}
Численное моделирование роста кремния в промышленных установках требует значительных вычислительных ресурсов, наиболее трудоемким является трехмерный расчет турбулентного течения расплава. 
Параллельные вычисления позволяют значительно снизить продолжительность цикла моделирования.

Эффективность параллельных вычислений характеризуется ускорением 
$S_n$  и полной эффективностью $E_n^{tot}$, определяемой как \cite{dri}
$S_n={T_s/T_p},\  E_n^{tot}={T_s/nT_p}$
где  $n$ - число процессоров, а  $T_s$ и $T_p$  --- время работы последовательного и параллельного алгоритмов, соответственно. 
Полная эффективность есть произведение трех факторов
 \cite{dur}: параллельной эффективности, численной эффективности и эффективности баланса нагрузки.
Первая определяется как 
$E^{par}_n={T_{comp}/(T_{comp} + T_{comm})},$
где  $T_{comp}$ - время вычислений и $T_{comm}$ --- время обмена сообщениями.

Решение уравнений в частных производных методами конечных разностей или конечных объемов на системах с централизованным управлением потоком данных ( Single Instruction Multiple Data - SIMD) 
требует разбиения расчетной сетки на подобласти, размер которых пропорционален мощности процессора
при минимизации обменов информацией между процессорами.
Это означает минимизацию целевой функции  
  $H = T_{comp} + T_{comm}$.
Первое слагаемое, очевидно, 
$T_{comp} = \max_q \left( N_q T_q\right)$,
где  $N_q$ --- число узлов (ячеек), приписанных процессору  $q$,
а $T_q$ --- время выполнения одной итерации  (производительность процессоров может различаться).
Время коммуникаций  $T_{comm}$ определяется  {\em коммуникационной матрицей} $C_{pq}$. $C_{pq}$  --- число узлов сетки, принадлежащих процессору $p$, которые необходимы процессору  $q$  для выполнения итерации.
Зависимость  $T_{comm}$  от $C_{pq}$ определяется архитектурой многопроцессорного  комплекса.  Для кластера рабочих станций  \cite{m1:levin}:
\begin{equation}
T_{comm} = \sum_{pq}\left( C_{pq}/b + \delta\left( C_{pq}\right) L\right),
\end{equation}
\label{m1:Tcomm}
где $L$ ---  латентность сети  (время, необходимое для начала обмена между парой процессоров) и  $b$ ---  {\em эффективная ширина пропускания} $b = 1/t_{vertex}$,
где $t_{vertex}$ - время, требуемое для передачи информации, относящейся к одной вершине.

Параллельная версия, использующая библиотеку  MPI, описана в   \cite{m1:paral}; существенным элементом является введение вспомогательного индекса для каждой ячейки многоблочной структурированной сетки, обеспечивающего непрерывность нумерации.
Расчеты проведены для роста кристаллов кремния 100 и 300 мм на двух кластерах рабочих станций, использующих Fast Ethernet и Myrinet; последний обеспечивает более высокую эффективность распараллеливания (ускорение составило 9.15 на системе из 20 процессоров).

\subsection{Образование}
Информационные технологии (ИТ) играют двоякую роль в инженерном образовании: с одной стороны, именно они в значительной степени формируют требования к выпускникам ВУЗов, с другой – предоставляют средства для создания образовательной среды нового поколения, существенно отличающейся  способом подачи знаний. Основными задачами образования являются  \cite{zp,nae,pra}:
многодисциплинарное образование и системный уровень подачи материала;
ориентация на практическое использование получаемых знаний, т. е. смещение акцента с "получения" знаний как набора сведений на развитие навыков их использования и генерацию нового знания, чтобы избежать ситуации "`выпусникник знает все, может ничего"';
адаптация средств обучения к личности обучаемого (его опыту, потребностям, стилю).

Компьютерно-Ориентированные Средства Обучения  (КОСО, в англоязычной литературе CBE – Computer Based Education) акцент делают на двух уникальных возможностях ИТ: 1) гипермедийные средства подачи материала и 2)
численное моделирование технического устройства или системы.

Первая дает статистически значимые сдвиги в эффективности обучения \cite{zyw} и становится de facto стандартом \cite{it}, но  «численное моделирование» часто сводится к анимации. Моделирование требует 1) наличия компьютерной модели реального технического устройства или системы; 2) предоставления обучаемому возможности экспериментировать с моделью.
Выполнение этих требований позволяет добавить слово "`инженерный"' в определение: Computer-Based {\em Engineering} Education  (CBEE). 
Пример такого ПО ("eduware"), основанного на моделировании сублимационного роста монокристаллов, описан в \cite{ed1,ed2}.

\section{Заключение}

Дан обзор тепломассобмена при росте монокристаллов. 
Рассмотрены особенности теплообмена теплопроводностью, конвекцией и излучением (с учетом полупрозрачности кристаллов и наличия диффузно и зеркально отражающих поверхностей) при росте из расплава и паровой фазы. Обсуждены вопросы характеризации качества монокристаллов, проблемы верификации моделей и оптимизации ростовых установок и режимов выращивания, особенности программного обеспечения, предназначенного для фундаментальных и прикладных исследований.

В заключение следует сделать ряд общих замечаний.
\begin{itemize}
\item{
Иногда утверждается, что зрелость программного обеспечения --- исследовательского или промышленного - определяется уровнем его верификации
\cite{98}. Это упрошение. Если требуется единственный параметр для характеризации практической  ценности моделирования, наилучший выбор ---
надежность предсказания \cite{h99}. Заметим, однако, что этот критерий характеризует не ПО как таковое, а моделирование, будучи зависимым от целей последнего. Очевидно, одни и те же расчеты можно рассматривать успешными, если
требовалось выявить некоторые тенденции, и неудачными, если искался  оптимальный размер элемента ростовой установки (например, экрана в установке для выращивания кремния методом Чохральского \cite{rs2,rs3,rs4,rs5,rs6}).
} 
\item{Джеффри Дерби как-то адаптировал известное высказывание Пикассо ("`Искусство --- ложь, которая заставляет понять правду"'): 
{\em Модель --- ложь, которая помогает понять правду} \cite{derby}.
Очевидно, нужно искать компромисс между полнотой модели и ее реализуемостью 
\cite{zhm04} (``модель описывает часть реальной системы, используя  {\em подобные}, но {\em более простые} структуры \cite{raabe}).
Правильное отношение к разработке модели можно заимствовать из письма Б. Паскаля от 14 декабря 1656 г.: "Я сделал письмо таким длинным, т. к. не имел времени сделать его короче". Создание модели заканчивается не тогда, когда  
учтены все физические явления 
(``общий подход --- включать все новые явления в моделирование'' \cite{duffar}), но когда модель {\em нельзя упростить, не разрушив ее} 
\footnote{
Так, к сожалению, иногда поступают, как, например, в недавней работе 
\cite{jin}, сообщающей о моделировании роста больших кристаллов кремния методом Чохральского в рамках стационарной осесимметричной модели, в которой пренебрегается эффектом Марангони, вращением тигля и кристалла,
а фазовая граница считается плоской; в работе  \cite{faiez} при моделировании роста гадолиниевого граната тем же методом боковая стенка тигля предполагалась изотермической, а нижняя - теплоизолированной. 
}.
Такую модель иногда называют минимальной.
}
	\item{Чтобы не утонуть в океане данных, следует помнить, что больше данных - не обязательно больше информации, больше информации - не обязательно больше знания, а больше знания - не обязательно больше мудрости - как полвека назад предупреждал Р. В. Хемминг в классической монографии 
"`Численные методы для научных работников и инженеров"' 
(McGraw-Hill, New York, 1962; Наука, М., 1968):
{\Large цель расчетов - понимание, а не числа}.
} 
\end{itemize}

Автор благодарен М. В. Богданову, А. О. Галюкову, И. А. Жмакину, В. В. Калаеву, С. Ю. Карпову, С. К. Кочугуеву, А. В. Кулику, Д. П. Луканину, Ю. Н. Макрову, М. Н. Немцевой,  Д. Х. Офенгейму,  М. С. Рамму, Э. А. Рудинскому, А. С. Сегалю, Е. М. Смирнову, В. С. Юфереву  за сотрудничество и полезные обсуждения.


\renewcommand{\baselinestretch}{1.}
{\small


\begin{thebibliography}{100}
\providecommand{\url}[1]{{#1}}
\providecommand{\urlprefix}{URL }
\expandafter\ifx\csname urlstyle\endcsname\relax
  \providecommand{\doi}[1]{DOI \discretionary{}{}{}#1}\else
  \providecommand{\doi}{DOI \discretionary{}{}{}\begingroup
  \urlstyle{rm}\Url}\fi

\bibitem{sapph}
Dobrovinskaya E, Lytvynov L A, Pishchik V \emph{Sapphire. Material,
  Manufacturing, Applications} (Springer, 2009)

\bibitem{srti}
Yoshimura J, Sakamoto T, Yamanaka J Jap. J. Appl. Phys. \textbf{40} 6536
  (2001)

\bibitem{2s}
Scheel Н J. \emph{Crystal Growth} \textbf{211}, 1 (2000)

\bibitem{ss1}
Карпентер Г  \emph{УФН} \textbf{X}, 689 (1930)

\bibitem{ss2}
Демьянец Л Н  \emph{УФН} \textbf{161} 71 (1991)

\bibitem{vegad}
Vegad M, Bhatt N  \emph{Procedia Technol.} \textbf{14} 438 (2014)

\bibitem{shimura}
Shimura F, in \emph{Springer Handbook of Electronic and Photonic Materials}
  (Eds. S Kasap, P Capper) (Springer, 2007) p. 255

\bibitem{bor}
Бородин В А  \emph{УФН} \textbf{170} 999 (2000)

\bibitem{ss4}
Shi Y H,  Dennis A R, Cardwell D A,
arXiv:1410.7973  (2014)

\bibitem{2}
Scheel H, in \emph{Crystal Growth Technology} (Eds  H Scheel, T Fukuda)
  (John Wiley \& Sons, 2003) p. 3

\bibitem{zuleh}
Zulencher W.
\newblock Hystorical overview of silicon crystal pulling development.
http://www.tf.uni-kiel.de/matwis/amat/elmat-en/articles/historic-review-cryst-growt.pdf

\bibitem{stri2}
Scheel H J.  \emph{Crystal Growth} \textbf{287} 214 (2006)

\bibitem{nacke}
Nacke B, Muizniecks A \emph{GAMM-Mitt.} \textbf{30} 113 (2007)

\bibitem{solar}
Lan C, Hsieh C, Hsu W, in \emph{Crystal Growth of Silicon for Solar Cells}
(Eds  K Nakajima, N Usami) (Berlin: Springer, 2009)  p. 25

\bibitem{lan}
Lan C  \emph{Chem. Engineer. Sci.} \textbf{59} 1437 (2004)

\bibitem{dash}
Дэш Э \emph{УФН} \textbf{LXXII} 495 (1960)

\bibitem{rud04}
Hurle D, Rudolph P \emph{J. Crystal Growth} \textbf{264} 550 (2004)

\bibitem{kyr2}
Chen C-H et al. \emph{J. Crystal Growth} \textbf{318} 162
  (2011)

\bibitem{prop2}
Jain S C et al.  \emph{J. Appl. Phys.}
  \textbf{87} 965 (2000)

\bibitem{epi}
Kasap S, Capper P (Eds), \emph{Springer Handbook of Electronic and Photonic
  Materials} (N.Y.: Springer, 2007)

\bibitem{orton}
Orton J W,  Foxon C T \emph{Rep. Prog. Phys.} \textbf{61} 1 (1998)

\bibitem{uv}
Nakamura S, Fasol G, Pearton S \emph{The Blue Laser Diode (2nd ed.)}
  (N.Y.: Springer, 2000)

\bibitem{zuk}
Zukauskas A,  Shur M S, Gaska R \emph{Introduction to Solid State Lighting}
  (N.Y.: Wiley, 2002)

\bibitem{schub}
Schubert F E \emph{Light-Emitting Diodes (2nd ed.)} (Cambride: CUP, 2006)

\bibitem{tran2}
Morkoc H \emph{Wide Band Gap Nitrides and Devices} (N.Y.: Springer, 1998)

\bibitem{peart}
Pearton S J, Ren F, Zhang A \emph{Mater. Sci. Eng.} \textbf{30} 55 (2000)

\bibitem{z10aln}
Avdeev O et al.,  in \emph{Crystal Growth Technology:
  Semiconductors and Dielectrics} (Eds P Capper, P Rudolph) (WILEY-VCH,
  2010)  p. 121


\bibitem{cry_is}
Schujman B S,  Schowalter L J
"Ga{N}-ready aluminum nitride substrates for cost-effective, very low
  dislocation density {I}{I}{I}-nitride {L}{E}{D}s"{\ },
Final Scientific/Technical Report   DE-FC26-08-NT01578 (Crystal IS, 2010)

\bibitem{cao07}
Cao X A et al.  \emph{J. Crystal Growth} \textbf{300} 382 (2007)

\bibitem{bu04}
Bu G et al \emph{Appl. Phys.  Lett.} \textbf{84} 4611 (2004)

\bibitem{epel_saw}
Epelbaum B M, Bickermann M, Winnacker A, in
\emph{Proc. 2nd Int. Symp.on Acoustic Wave Devices for Future Mobile Comm.
  Systems, Chiba} (2004) p.157

\bibitem{fritze}
Fritze H \emph{J. Electroceram.} \textbf{26} 122 (2011)

\bibitem{cleland}
Cleland A N, Pophristic M, Ferguson I \emph{Appl. Phys. Lett.} \textbf{79} 2070
  (2001)

\bibitem{sub}
Kukushkin S A et al. \emph{Rev. Adv. Mater. Sci} \textbf{17} 1 (2008)

\bibitem{speck2}
Speck J S  \emph{Mater. Sci. Forum} \textbf{353-356} 769 (2001)

\bibitem{paskova10}
Paskova T, Hanser D A, Evans K R \emph{Proc. IEEE} \textbf{98} 1324 (2010)

\bibitem{bul12}
Bulashevich K et al., in
  \emph{Light-Emitting Diodes and Optoelectronics: New Research} (NOVA SCIENCE
  PUBLISHERS, INC, 2012), p. 231

\bibitem{spr2}
Bogdanov M V et al. \emph{Semicond. Sci. Technol} \textbf{23} 125023 (2008)

\bibitem{zhm11}
Zhmakin A I  \emph{Phys. Rep.} \textbf{498} 189 (2011)

\bibitem{2in}
Russel P, in \emph{CS MANTECH Conf., Vancouver} (2006) p.231

\bibitem{sitar}
Dalmau R, Sitar Z, in  \emph{Encyclopedia of Materials: Science and Technology}
(Eds. K H J Buschow  et al.) (Elsevier, 2005) p. 9

\bibitem{denis}
Denis A, Goglio G, Demazeau G \emph{Mater. Sci. Eng.} \textbf{R50} 167 (2006)

\bibitem{ipap}
Schowalter L J et al. \emph{IPAP Conf. Series} \textbf{4}  38
  (2004)

\bibitem{am}
Ehrentraut D, Fukuda T, in \emph{Frontiers in Materials Research} (Eds 
  Y Fujikawa, K Nakajima, T Sakurai) (Springer, 2008) p. 111

\bibitem{g1}
		Krukowski S \emph{Diamond Relat. Mat.} \textbf{6} 1515 (1997)

\bibitem{cg}
Ohtani N et al.,  in \emph{Wide Bandgap Semiconductors.
  Fundamental Properties and Modern Photonic and Electronic Devices} (Eds
  K Nakahashi, A Yoshikawa, A Sandhu) (Springer, 2007) p. 329

\bibitem{utsumi}
Utsumi W et al.  \emph{Nat.  Mater.} \textbf{2} 735 (2003)

\bibitem{g3}
Nord J et al.  \emph{J. Phys.: Condens Matter} \textbf{15}
  5649 (2003)

\bibitem{kawa}
Kawahara M et al.  \emph{J. Ceramic Process. Res.} \textbf{6} 146 (2005)

\bibitem{por}
Porowski S, Grzegory I \emph{J. Crystal Growth} \textbf{178} 174 (1997)

\bibitem{g4}
Grzegory I \emph{J. Phys.: Condens Matter} \textbf{13} 6875 (2001)

\bibitem{bock}
Bockowski M \emph{Cryst. Res. Technol.} \textbf{36} 771 (2001)

\bibitem{bel1}
Belousov A "High pressure crystal growth, thermodynamics and physical properties
  of {A}l$_x${G}a$_{1-x}${N} semiconductors", Dr. Sci. thesis (Z\"urich: Eidgen\"ossische Technische Hochschule, 2010)

\bibitem{5}
Purdy A, Case S, Muratore N   \emph{J. Crystal Growth} \textbf{252} 136 (2003)

\bibitem{13}
Yoshikawa A et al. \emph{J. Crystal Growth}   \textbf{260} 67 (2004)

\bibitem{g19}
Aoki M et al. \emph{J. Crystal Growth} \textbf{246}   133 (2002)

\bibitem{g20}
Song Y et al. \emph{J. Crystal Growth} \textbf{247} 275
  (2003)

\bibitem{ameth}
Carter C B,  Norton M G  \emph{Ceramic Materials} (Springer N.Y., 2007)

\bibitem{quartz}
Iwasaki F, Iwasaki H \emph{J. Crystal Growth} \textbf{237-239} 820 (2002)

\bibitem{cal13}
Ehrentraut D et al. \emph{Prog. Cryst.  Growth Charact. Mater.} \textbf{52} 280 (2006)

\bibitem{cal9}
Nause J, Nemeth B  \emph{Semicond. Sci. Technol.} \textbf{20} S45 (2005)

\bibitem{calla}
Callahan M J, Q.S. Chen Q S, in 
\emph{Springer Handbook of Crystal Growth} (Eds  Dhanaraj  G et al.) (Springer, 2010) p. 655

\bibitem{fukuda}
	Fukuda T, Ehrebtraut D \emph{J. Crystal Growth}  \textbf{305} 304 (2007)

\bibitem{s6}
Karpov S Yu et al.  \emph{phys. stat. sol. (a)} \textbf{176} 435 (1999)

\bibitem{brink}
Brinkman A W , Carles J \emph{Progr. Crystal Growth Character. Mater.} 169  (1998)

\bibitem{tai}
Tairov Y, Tsvetkov V \emph{J. Crystal Growth} \textbf{43} 209 (1978)

\bibitem{sitar10}
Dalmau R, Sitar Z,  in \emph{Springer Handbook of Crystal Growth} (Eds G Dhanaraj et al.) (Spriner, 2010) p. 821

\bibitem{z11}
Avdeev O et al., in  \emph{Comprehensive Semiconductor   Science and Technology} (Eds P Bhattacharya, R Fornari, H  Kamimura) (ELSEVIER SCIENCE, 2011)  \textbf{3}  p. 282

\bibitem{z12}
Avdeev O et al. in \emph{Modern Aspects of Bulk Crystal and Thin Film
  Preparation} (Eds Kolesnikov N, Borisenko E) (INTECH, 2012), p. 213

\bibitem{ro06}
Rojo C J et al.  \emph{Proc. SPIE} \textbf{6122} 61220Q1
  (2006)

\bibitem{fib}
Bao H Q et al. \emph{Appl. Phys. A - Mater.   Sci. \& Process.} \textbf{94} 173 (2009)

\bibitem{titan}
Du L et al. \emph{J. Mater. Sci.: Mater. Electron.}   \textbf{21} 78 (2010)

\bibitem{kallinger}
Kallinger B et al. \emph{Cryst. Res. Techn.} \textbf{43} 14 (2008)

\bibitem{freitas10}
Freitas J A \emph{J. Phys. D: Appl. Phys.} \textbf{43} 073301 (2010)

\bibitem{tgt}
Xu J et al. \emph{J. Crystal Growth} \textbf{193}   123 (1998)

\bibitem{ligo}
Zhou G et al. \emph{Material Lett.}   \textbf{60} 901 (2006)

\bibitem{ligo1}
Barish B  et~al. \emph{IEEE Trans. Nuclear Sci.}   \textbf{49} 1233 (2002)

\bibitem{bagd}
Багдасаров Х С,  Хаимов-Мальков В Я \emph{УФН} \textbf{102} 324 (1970)

\bibitem{zeng}
Zeng Z et al. \emph{Mater. Trans.} \textbf{45} 1515 (2004)

\bibitem{derby1}
Derby J et al.,  in \emph{Perspective on Inorganic, Organic and Biological Crystal
  Growth: From Fundamentals to Applications} (Eds Sowronski M, DeYoreo Y, Wang C) (2007)  p. 139

\bibitem{mpc2}
Ng J, Dubljevich S  \emph{J. Process Control} \textbf{21} 1361 (2011)

\bibitem{rud07}
Rudolph P \emph{Function. Mater.} \textbf{14} 411 (2007)

\bibitem{jang} 
Jang J, Kye U,  Kang C J,  arXiv:1408.4261  (2014)

\bibitem{chern1}
Чернов А А \emph{УФН} \textbf{73}, 277  (1961)

\bibitem{chern2}
Чернов А А  \emph{УФН} \textbf{153}, 678  (1987)

\bibitem{weeks}
Weeks J, Gilmer G, in \emph{Advances in Chemical Physics} \textbf{40} (Ed Prigogine I, Rice C)  (2007)  p. 157

\bibitem{raffi}
Raffi-Tabar H, Chiraz A \emph{Phys. Rep.} \textbf{365} 145 (2002)

\bibitem{10}
Shyy W Udaykumar H, in \emph{Computational Analysis of Convective Heat
  Transfer} (Eds   Sunden B,  Comini G) (Southampton: WIT,  2000)  p.   141

\bibitem{11}
Kaempfer T, Rappaz M \emph{Model. Simul. Mater. Sci. Eng.} \textbf{11} 575 (2003)

\bibitem{dend}
Александров Д В, Галенко П К \emph{УФН} \textbf{184} 833 (2014)

\bibitem{3}
	Chiu C et al. \emph{J. Mat. Res.} \textbf{9} 2066 (1994)

\bibitem{vartak}
Vartak B, Yeckel A, Derby J \emph{J. Crystal Growth} \textbf{281} 391 (2005)

\bibitem{levy}
Levy Y, J.N. Onuchic J N \emph{Annu. Rev. Biophys. Biomol. Struct.} \textbf{35} 389
  (2006)

\bibitem{rand}
Rand R P  \emph{Phil. Trans. R. Soc. Lond. B} \textbf{359} 1277 (2004)

\bibitem{smolin} 
Smolin N "A simulation analysis of protein hydration"  Dr. rer. nat. thesis (Universit\"at Dortmund, 2006)

\bibitem{4}
Garbuzov D et al.   \emph{Crystal Growth} \textbf{110} 955 (1991)

\bibitem{abba}
Abbaso\^glu S \emph{Turk. J. Elec. Eng. \& Comp. Sci.} \textbf{20} 1343 (2012)

\bibitem{crn08}
Crnogorac N et al. \emph{Cryst. Res. Technol.}   \textbf{43} 606 (2008)

\bibitem{1}
Meyappan M (ed.) \emph{Computational Modeling in Semiconductor Processing}
  (The Artech House, 1994)

\bibitem{96}
Ofengeim D,  Zhmakin A \emph{Lect. Notes Comput. Sci.} \textbf{2657} 3 (2003)

\bibitem{zhm04}
Zhmakin A I, in 
\emph{CD-ROM Proc. Third Int. Symp.  Adv. Comput.  Heat Transfer} (Begell, 2004)
Keynote lecture

\bibitem{bogd}
Bogdanov M V, Ofengeim D K, . Zhmakin A I \emph{Centr. Europ. J. Phys.} \textbf{2}
  183 (2004)

\bibitem{14}
Bogdanov M et al. \emph{Crystal   Res.Techn.} \textbf{38} 237 (2003)

\bibitem{vr1}
Bogdanov M V et al. \emph{J. Crystal   Growth} \textbf{225} 307 (2001)

\bibitem{quasi1}
Basu B et al. \emph{ J. Crystal Growth} \textbf{230} 148
  (2001)

\bibitem{41}
Makarov Yu, Zhmakin A \emph{J. Crystal Growth} \textbf{94} 537 (1989)

\bibitem{kulk}
Kulkarni M, Voronkov V, R.~Falster R \emph{J. Electrochem. Soc.} \textbf{151} G663
  (2004)

\bibitem{tal}
Talanin V, Talanin I, Ustimenko N \emph{Sci. \& Technol.} \textbf{2} 130 (2012)

\bibitem{ammon}
Ammon W  in [20]  p. 101

\bibitem{transient}
Klein O, P.~Philip P, Sprekels J \emph{Interfaces and Free Boundaries} \textbf{6}
  295 (2004)

\bibitem{tr1}
Wang S et al. \emph{Int. J. Heat Mass   Transfer} \textbf{84} 370 (2015)

\bibitem{7}
Guide for the verification and validation of computational fluid dynamics
  simulations.
\newblock AIAA G-077-1998 (1998)

\bibitem{raabe}
Raabe D  \emph{Adv. Mater.} \textbf{14} 639 (2002)

\bibitem{chen08}
Chen X J et al. \emph{J. Crystal Growth}   \textbf{310} 1810 (2008)

\bibitem{trial}
Dupret F et al., in  \emph{The 19th American Conf. on Crystal Growth and Epitaxy, 
  2013, Keystone, Colorado}  http://science24.com/paper/29805

\bibitem{8}
Luft J et al. \emph{J. Crystal Growth} \textbf{196} 447 (1999)

\bibitem{kurz}
Kurz M, M\"uller G \emph{J. Crystal Growth} \textbf{208} 341 (2000)

\bibitem{9}
Тихонов А Н, Арсенин В Я  \emph{Методы решения некорректных задач} (М: Наука, 1974)


\bibitem{zhm90}
Жмакин А И, Макаров Ю Н  \emph{ЖТФ} \textbf{60} 37 (1990)

\bibitem{ipa}
Жмакин А И, Ипатова И П, Макаров Ю Н \emph{ЖТФ} \textbf{56} 1700
  (1986)

\bibitem{zhm90_m}
Жмакин А И, Макаров Ю Н \emph{Матем. моделирование}   \textbf{2}(7) (1990)

\bibitem{37}
Evstratov I et al. \emph{J. Crystal   Growth} \textbf{230} 22 (2001)

\bibitem{z8si}
Kalaev V, Makarov Yu, Zhmakin A, in \emph{Crystal Growth Technology. From
  Fundamentals and Simulation to Large-scale Production} (Eds H J Scheel,
  P Capper) (Weinheim: Wiley-VCH, 2008) p. 173

\bibitem{z10}
Demina S et al., in \emph{Crystal Growth  Technology: Semiconductors and Dielectrics} (Eds P Capper, R Rudolph)
  (Wiley-VCH, 2010), p. 213

\bibitem{mart14}
Martinez A M et al. \emph{J. Mater. Phys. Chem.}  \textbf{2} 9 (2014)

\bibitem{38}
Kalaev V, Zhmakin A. in \emph{Proc. of 9th European Turbulence Conference, Southampton} (2002) p 207 

\bibitem{kak13}
Kakimoto K \emph{Acta Physica Polonica A} \textbf{124} 227 (2013)

\bibitem{mar14}
Martinez A et al. \emph{ J. Mater. Phys. Chem.}  \textbf{2} 9 (2014)

\bibitem{12}
Song Y et al. \emph{J. Crystal Growth}   \textbf{260} 327 (2004)

\bibitem{13a}
Chen Q S, Prasad V, Hu W \emph{J. Crystal Growth} \textbf{258} 181 (2003)

\bibitem{15}
Dhanaraj  G et al.  in \emph{Crystal Growth Technology} (Eds Byrappa K, Ohashi T (William Andrew Publ., 2003) p. 181

\bibitem{16}
Kitanin E et al. \emph{Mater. Sci. Eng.} \textbf{B55} 174
  (1998)

\bibitem{19}
Hurle D (ed.) \emph{Handbook of Crystal Growth. Bulk Crystal Growth}
  (North-Holland, 1994)

\bibitem{dupret94}
Dupret F,  den Bogaert N V, in  [133] \textbf{2} 877

\bibitem{21}
Lan C, Tu C  \emph{J. Crystal Growth} \textbf{226} 406 (2002)

\bibitem{22}
Ozawa T et al. \emph{J. Crystal Growth}   \textbf{237-239} 1692 (2002)

\bibitem{23}
Kalaev V, Evstratov I, Makarov Y \emph{J. Crystal Growth} \textbf{249} 87 (2003)

\bibitem{24}
Muehlbauer  A, Muiznieks A, Raming G  \emph{Cryst. Res. Technol.} \textbf{34} 217
  (1999)

\bibitem{minak}
Minakuch H et al. \emph{J. Advanc.  Res. Phys.} \textbf{3} 011201 (2012)

\bibitem{sige1}
Ashburn P, Bagnall D in [20]  p. 481

\bibitem{lpd}
Usami N,  in \emph{Silicon-Germanium (SiGe) Nanostructures - Production,
  Properties and Applications in Electronics} (Eds Shiraki Y, Usami N)
  (Cambridge: Woodhead Publishing,  2011) p. 72

\bibitem{lpd1}
Sekhom M, Armour N, Dost S \emph{FDMP} \textbf{9} 331 (2013)

\bibitem{bolkh}
Болховитянов Ю Б, Пчеляков О П, Чикичев СИ \emph{УФН}
  \textbf{171} 689 (2001)

\bibitem{sige}
Dario O, H.~Sicim H, Balicki E, \emph{J. Crystal Growth} \textbf{318} 1057 (2011)

\bibitem{armour}
Armour N, Dost S  \emph{Acta Physica Polonica A} \textbf{124} 198 (2013)

\bibitem{25}
Yuferev V et al. \emph{Theoret.  Comput. Fluid Dynamics} \textbf{12} 53 (1998)

\bibitem{25a}
Yuferev V et al.  \emph{J. Crystal   Growth} \textbf{180} 578 (1997)

\bibitem{27}
Fedyushkin A et al. in \emph{{P}roc. 2nd {P}an
  {P}acific {B}asin {W}orkshop on {M}icrogravity {S}ciences}, paper cg-1065
  (2002)

\bibitem{pol}
Полежаев В И и др.  \emph{Математическое моделирование конвективного тепломассообмена} (М: Наука,  1987)

\bibitem{6}
Li H, E.~Evans E,  Wang G X \emph{Crystal Growth} \textbf{256} 146 (2003)

\bibitem{28}
Vizman D , Grabner O, M\"uller G \emph{J. Cryst. Growth} \textbf{233} 687 (2001)

\bibitem{Langlois_77}
Langlois W \emph{J. Crystal Growth} \textbf{42} 386 (1977)

\bibitem{Samarskii_86}
Самарский А А , Попов Ю П, Мажорова О С \emph{
Математическое моделирование. Получение монокристаллов и полупроводниковых структур}
(М: Наука, 1986) 

\bibitem{Brown_88}
Brown R  \emph{AIChE J.} \textbf{34} 881 (1988)

\bibitem{Muller_88}
M\"uller G  in \emph{Crystals} \textbf{12} (Ed. Freyhardt  H) (Berlin, Heidelberg: Springer, 1988)

\bibitem{BottaroZebib}
Bottaro A, Zebib A \emph{J. Crystal Growth} \textbf{97} 50 (1989)

\bibitem{Seidl}
Seidl A et al. \emph{J. Crystal Growth} \textbf{137}   326 (1994)

\bibitem{Kakimoto}
Kakimoto K et al. \emph{J. Crystal Growth}   \textbf{139} 197 (1994)

\bibitem{Hurle}
Hurle D, Cockayane B, in  [133]  \textbf{2}  p. 99

\bibitem{Ristorcelli_a}
Ristorcelli J, Lumley J \emph{J. Crystal Growth} \textbf{116} 447 (1992)

\bibitem{LaunderSpalding}
Launder B, Spalding D \emph{Comput. Methods Appl. Mech. Eng.} \textbf{3} 269
  (1974)

\bibitem{Kobayashi90}
Kobayashi S \emph{J. Crystal Growth} \textbf{99} 692 (1990)

\bibitem{KinneyBrown}
Kinney T, Brown R \emph{J. Crystal Growth} \textbf{132} 531 (1993)

\bibitem{Ristorcelli_b}
Ristorcelli J, Lumley J \emph{J. Mater. Process. Manufact. Sci.} \textbf{1} 69  (1992)

\bibitem{farge}
Farge M et al. \emph{Proc. IEEE} \textbf{84} 639
  (1996)

\bibitem{29}
Kalaev V, Zhmakin A, Smirnov E \emph{J. of Turbulence} \textbf{3} 013 (2002)

\bibitem{30}
Lipchin A, Brown R \emph{J. Crystal Growth} \textbf{216} 192 (2000)

\bibitem{corr}
Wallace J   \emph{Theor. Appl. Mech. Lett.} \textbf{4} 022003 (2014)

\bibitem{moin}
Moin P, Manesh K \emph{Annu. Rev. Fluid Mech.} \textbf{30} 539 (1998)

\bibitem{antt}
Anttila O \emph{ECS Trans.} \textbf{3} 3 (2006)

\bibitem{taylor}
Landahl M, Mollo-Christensen E, \emph{Turbulence and Random Processes in
  Fluid Mechanics, 2nd ed.} (Cambridge, 1992)

\bibitem{32}
Renner C et al., arXiv:physics/0109052

\bibitem{34}
Yokokawa M et al. 16.4 {T}{F}lops direct numerical simulation of turbulence by a
  {F}ourier spectral method on the {E}arth {S}imulator.
 http://www.sc-2002.org/paperpdfs/pap.pap273.pdf

\bibitem{34a}
Ishihara T, Goto T, Kaneda Y \emph{Annu. Rev. Fluid Mech.} \textbf{41}   165 (2009)

\bibitem{35}
Egorov Y et al. \emph{Mat. Res. Soc.  Proc.} \textbf{490} 181 (1998)

\bibitem{36}
Evstratov I et al. \emph{Microelectr. Eng.} \textbf{56} 139 (2001)

\bibitem{organ}
Organ N et al. \emph{J. Crystal Growth} \textbf{82} 481 (1987)

\bibitem{quasi}
Enger S et al. \emph{J. Crystal Growth} \textbf{219} 144   (2000)

\bibitem{farge92}
Farge M \emph{Annu. Rev.  Fluid Mech.} \textbf{24}, 395 (1992)

\bibitem{schneid}
Schneider K,  Kevlahan N R, Farge M \emph{Theoret. Comput. Fluid Dynamics}   \textbf{9} 191 (1997)

\bibitem{liu12}
Liu L, Liu X, Wang Y \emph{Int. J. Heat Mass Transfer} \textbf{55} 53 (2012)

\bibitem{38a}
Kalaev V, Zhmakin A, in \emph{Engineering Turbulence Modelling and
  Experiments} \textbf{5} (Eds W Rodi, N Fueyo) (Elsevier, 2002) p.   337

\bibitem{ev02}
Evstratov I  et al.  \emph{J. Crystal Growth}   \textbf{237-239} 1757 (2002)

\bibitem{zeyt83}
Зейтунян Р Х  ПМТФ (2) 53 (1983)

\bibitem{roach}
Роуч П \emph{Вычислительная гидродинамика}   (М: Мир, 1980)

\bibitem{lapin}
Лапин Ю В, Стрелец М Х \emph{Внутренние течения  газовых смесей} (М: Наука, 1989)

\bibitem{zhm01}
Жмакин А И др., в сб. \emph{Вопросы   математической физики и прикладной
  математики}(СПб: ФТИ  им.Иоффе РАН, 2001) с. 208

\bibitem{fed}
Федорченко А Т \emph{ЖВММФ} \textbf{21} 1216 (1981)

\bibitem{40}
Жмакин А И, Макаров Ю Н \emph{Докл. АН СССР} \textbf{280} 827 (1980)

\bibitem{zey}
Zeytounian R K \emph{Theory and Applications of Viscous Fluid Flows}
  (Berlin: Springer,  2003)

\bibitem{ober}
Oberbeck A \emph{Ann. Phys. Chem.} {\bf 7} 271  (1879)

\bibitem{42}
Egorov Y,  Zhmakin A  \emph{Comput. Mater. Sci.} \textbf{11} 204 (1998)

\bibitem{43}
Kleijn C \emph{Thin Solid Films} \textbf{365} 294 (2000)

\bibitem{vor}
Vorob'ev A et al. \emph{J. Crystal Growth} \textbf{211} 343 (2000)

\bibitem{44}
Vorob'ev A et al. \emph{Comput. Mat. Sci.} \textbf{24} 520 (2002)

\bibitem{45}
Rivas D  \emph{J. Crystal Growth} \textbf{262} 48 (2004)

\bibitem{47}
Dupret F et al. \emph{Int. J. Heat Mass  Transfer} \textbf{33} 1849 (1990)

\bibitem{ern}
Ern A, Guermond J L \emph{Int. J. Numer. Meth. Eng.} \textbf{61} 559 (2004)

\bibitem{vf1}
Kumar A, Banerjee J, Muralidhar \emph{J. Sci. Industr. Res.} \textbf{61}  607   (2002)

\bibitem{ruk}
Rukolaine S et al.,  in \emph{Proc. 4th Int.
  Conf. Single Crystal Growth and Heat \& Mass Transfer} (2001) p. 669

\bibitem{z8}
Kalaev V et al.,  in \emph{Crystal Growth   Technology. From Fundamentals and Simulation to Large-scale Production}
(Eds H J Scheel, P Capper) (Weinheim: Wiley-VCH, 2008) p. 205

\bibitem{48}
Brandon S, Derby J   \emph{J. Crystal Growth} \textbf{110} 481 (1991)

\bibitem{49}
Xiao Q, Derby J \emph{J. Crystal Growth} \textbf{128} 188 (1993)

\bibitem{m2:yuf87}
Yuferev V, Vasiliev M  \emph{J. Crystal Growth} \textbf{82} 31 (1987)

\bibitem{50}
Yuferev V et al. \emph{J. Crystal Growth} \textbf{253} 383 (2003)

\bibitem{51}
Linhart J \emph{Cryst. Res. Technol.} \textbf{37} 849 (2002)

\bibitem{52}
Jing C et al.  \emph{J. Crystal Growth} \textbf{259} 367 (2003)

\bibitem{53}
Thomas P et al.  \emph{J. Crystal Growth}   \textbf{99} 133 (1989)

\bibitem{54}
Molchanov A et al. \emph{Cryst. Res. Technol.} \textbf{33} 207 (1998)

\bibitem{55}
Nicoara I et al. \emph{Cryst. Res. Technol.}  \textbf{33} 207 (1998)

\bibitem{56}
Nunes E et al.  \emph{J. Crystal Growth} \textbf{236} 596
  (2002)

\bibitem{57}
Rukolaine S et al. \emph{J. Quant. Spectr. Radiat.  Transfer} \textbf{73} 205 (2002)

\bibitem{m2:y3}
Xiao Q, Derby J \emph{J. Crystal Growth} \textbf{139} 147 (1994)

\bibitem{m2:y13}
Tsukada T et al. \emph{Int. J. Heat Mass Transfer}   \textbf{38} 2707 (1995)

\bibitem{m2:y4}
Kobayashi M et al. \emph{J. Crystal Growth}   \textbf{235} 258 (2002)

\bibitem{m2:y15}
Kobayashi M, Tsukado T,  Hozawa M  \emph{J. Crystal Growth}  \textbf{241} 241 (2002)

\bibitem{m2:y16}
Schwabe D,  Sumathi R R, Wilke H, \emph{J. Crystal Growth} \textbf{265} 440 (2004)

\bibitem{galaz}
Galazka Z, Schwabe D, Wilke H, \emph{Cryst. Res. Technol.} \textbf{38} 859 (2003)

\bibitem{ban}
Banerjee J, K.~Muralidhar K \emph{J. Crystal Growth} \textbf{286} 350 (2006)

\bibitem{z9}
Zhmakin A \emph{Fundamentals of Cryobiology. Physical phenomena and
  mathematical models. Springer Series “Biological and Medical Physics”}
  (Berlin: Springer, 2009)

\bibitem{74}
Golyshev V, Gonik M, Tsvetovsky V J. Crystal Growth \textbf{237-239} 735
  (2002)

\bibitem{shock}
Luo S N et al. \emph{Phys. Rev. B}   \textbf{69} 134206 (2003)

\bibitem{shock1}
Kuksin A  \emph{Comput. Phys. Comm.}   \textbf{177} 34 (2007)

\bibitem{pre}
Wettlaufer J S,  Worster M G,  \emph{Annu. Rev. Fluid Mech.} \textbf{38} 427 (2006)

\bibitem{m2:czsi}
Guo Z, Maruyama S, Tsukado T,  \emph{Numer. Heat Transf. Part A} \textbf{32} 595
  (1997)

\bibitem{m2:mam}
Мамедов В, в   сб. \emph{Вопросы   математической физики и прикладной
  математики}(СПб: ФТИ  им.Иоффе РАН, 2005) с. 198

\bibitem{m2:y_m}
Мамедов В М, Юферев В С  \emph{ТВТ} \textbf{44}  568 (2006)

\bibitem{m2:y22}
Мамедов В М, Руколайне С А \emph{Матем. моделирование} \textbf{16}  15 (2004)

\bibitem{how}
Howell J, Perlmutter M \emph{J. Heat Transfer} \textbf{86} 116 (1964)

\bibitem{ker}
Kersch A, Morokoff W \emph{Transport Simulation in Microelectronics}
  (Basel--Boston--Berlin: Birkh\"auser, 1995)

\bibitem{58}
Maruyama S, Aihira T, \emph{J. Heat Transfer} \textbf{119} 129 (1997)

\bibitem{59}
Kochuguev S et al. \emph{CFD J.} \textbf{II-33} 440   (2001)

\bibitem{koch1}
Kochuguev S et al. \emph{Proc. 16th IMACS World Congress, Lausanne} (2000)

\bibitem{koch}
Kochuguev S, Ofengeim D, Zhmakin A, in \emph{Proc. 3rd European Conf. on
  Num. Math. Adv. Appl.} (Eds P Neittaanm\"aki,
  T Tiihonen, P Tarvainen) (Singapore: World Scientific, 2000) p. 579

\bibitem{60}
Kochuguev S et al., in \emph{3rd IMACS Seminar on Monte Carlo Methods, Salzburg} (2001) p. 198

\bibitem{interfa}
Davidchack R L,  Laird B B \emph{Phys. Rev. Lett.} \textbf{85} 4751 (2000)

\bibitem{68}
Bene\^s  M \emph{Acta Math. Univ. Comenianae} \textbf{LXX} 123 (2001)

\bibitem{solid}
Hu H, Argyropoulos A \emph{Modell. Simul. Mater. Sci. Eng.}  (1996)

\bibitem{conv}
Cannon J R, DiBenedetto E,  Knightly G H  \emph{Arch. Ration. Mech. Anal.}   \textbf{73} 79 (1980)

\bibitem{nedoma}
Nedoma J \emph{J. Comput. Appl. Math} \textbf{84} 45 (1997)

\bibitem{griebel}
Griebel M, Merz W, Neunhoeffer T \emph{Comput. Visual. Sci.} \textbf{1} 201
  (1999)

\bibitem{rosenb}
Rosenberger F \emph{Fundamentals of Crystal Growth I. Macroscopic Equilibrium
  and Transport Concepts. Springer Series in Solid-State Sciences}
  (Berlin Heidelberg New York: Springer-Verlag, 1979)

\bibitem{69}
Barat C et al. \emph{Crystal Res. Technol.}   \textbf{34} 449 (1999)

\bibitem{70}
Dutta P, Ostrogorski A \emph{J. Crystal Growth} \textbf{194} 1 (1998)

\bibitem{71}
Reid D et al. \emph{J. Crystal Growth}   \textbf{174} 250 (1997)

\bibitem{72}
Coriel S, McFadde G \emph{J. Crystal Growth} \textbf{237-239} 8 (2002)

\bibitem{73}
Fujiwara K et al. \emph{J. Crystal Growth} \textbf{243} 275 (2002)

\bibitem{75}
M\"uller G  \emph{J. Crystal Growth} \textbf{237-239} 1628 (2002)

\bibitem{76}
Borovlev Y et al.  \emph{J. Crystal  Growth} \textbf{229} 305 (2001)

\bibitem{77}
Virozub A, Brandon S, \emph{Model. Simul. Mater. Sci. Eng.} \textbf{10} 57 (2002)

\bibitem{epi1}
Kolesnikov N (ed.), \emph{Modern Aspects of Bulk Crystal and Thin Film
  Preparation} (InTech, 2012)

\bibitem{sic_e}
Spencer M.
\newblock Si{C} growth technology in {E}urope.
\newblock http://www.wtec.org/loyola/ttec/hte-e/report/hte-chap4.pdf

\bibitem{137}
Bogdanov M V et al. \emph{Mat. Sci. Forum} \textbf{353-356} 57 (2001)

\bibitem{wolfson}
Wolfson A A, Mokhov E N \emph{Semicond.} \textbf{44} 1383 (2010)

\bibitem{m1}
Slack G A,  McNelly T F \emph{J. Crystal Growth} \textbf{34} 263 (1977)

\bibitem{m2}
Rojo J S et al.  \emph{Mat. Res. Soc. Symp. Proc.} \textbf{722} K1.1.1 (2002)

\bibitem{m3}
Raghothamachar B et al. \emph{J. Crystal  Growth} \textbf{250} 244 (2003)

\bibitem{hart}
Hartmann C et al.  \emph{J. Crystal Growth}   \textbf{310} 930 (2008)

\bibitem{miyan}
Miyanaga M et al.  \emph{SEI  Techn. Rev.} \textbf{63} 22 (2006)

\bibitem{lu06}
Lu P "Sublimation growth of {A}l{N} bulk crystals and high-speed {C}{V}{D}
  growth of {S}i{C} epilayers, and their characterization",  Ph. D. thesis (Kansas State Univ., 2006)

\bibitem{lu08}
Lu P et al.  \emph{Appl. Phys.  Lett.} \textbf{93} 131922 (2008)

\bibitem{mokh03}
Mokhov E N et~al. \emph{Mat. Sci. Forum} \textbf{433-436} 979 (2002)

\bibitem{lee07}
Lee C E et al.  \emph{IEEE Photonics Technol. Lett.} \textbf{19} 1200 (2007)

\bibitem{nov04}
	Noveski V et al.  \emph{MRS Internet J. Nitride Semicond. Res.} \textbf{9} 2 (2004)

\bibitem{bond07}
Bondokov R T et al. \emph{Mater. Res. Soc.  Symp. Proc.} \textbf{955} I03 (2007)

\bibitem{wang}
Wang X, Cai D, Zhang H \emph{AIAA 2006-3270} (2006)

\bibitem{rojo}
de~Almeida V F,  Rojo J C  Oak Ridge Nat. Lab., Tech. Rep. ORNL/TM-2002/64 (Oak Ridge Nat. Lab., 2002)

\bibitem{cai}
Cai D et al.  \emph{AIAA 2006-3825} (2006)

\bibitem{yazdi}
Yazdi G R, Syv\"aj\"arvi M, Yakimova R \emph{Phys. Scr.} \textbf{T126} 127 (2006)

\bibitem{sitar04}
Sitar Z et al.  \emph{IPAP Conf. Series} \textbf{4} 41 (2004)

\bibitem{def}
Klapper H, in \emph{Springer Handbook of Crystal Growth} (Eds G Dhanaraj,
  K Byrappa, V Prasad, M Dudley)  (Springer, 2010)

\bibitem{79}
Zhmakin I et al. \emph{Diamond  Relat. Mat.} \textbf{9} 446 (2000)

\bibitem{oxford}
Kochuguev S K et al., in
  \emph{Numerical Methods for Fluid Dynamics VII} (Ed  M Baines) (Oxford: OUP, 2001) p. 363

\bibitem{edgar}
Shi Y et al. \emph{MRS   Internet J. Nitride semicond. Res.} \textbf{6} 7 (2001)

\bibitem{yak}
Yakimova R et al. \emph{J. Crystal Growth} \textbf{281} 81 (2005)

\bibitem{s1}
Dryburgh P M  \emph{J. Crystal Growth} \textbf{125} 65 (1992)

\bibitem{s4}
Noveski V et al. \emph{J. Crystal Growth}   \textbf{266} 369 (2004)

\bibitem{s5}
Wu B et al. \emph{J. Crystal Growth} \textbf{266} 303 (2004)

\bibitem{karp01}
Karpov S Y et al. \emph{phys. stat. sol. (a)}   \textbf{188} 763 (2001)

\bibitem{s8}
Segal A S et al.  \emph{J. Crystal Growth} \textbf{211} 68 (2000)

\bibitem{s9}
Segal A S et al \emph{J.  Crystal Growth}  \textbf{270} 384 (2004)

\bibitem{s10}
Dreger L H, Dadape V V,  Margrave J L \emph{J. Phys. Chem.} \textbf{66} 1556   (1962)

\bibitem{s11}
Fan Z Y, Newman N \emph{Mat. Sci. Eng. B} \textbf{87} 244 (2001)

\bibitem{143}
Virtual reactor: Sublimation bulk crystal growth simulator, version 4.1, theory
  manual (Saint-Petersburg: Saint-Petersburg, 2003)

\bibitem{mf1}
von Ammon W et al., in \emph{The 15$^{th}$ Riga and 6$^{th}$ PAMIR Conf. on Fundamental and
  Applied MHD, Invited Lecture} 
  http://pamir.sal.lv/2005/cd/vol.I/riga-pamir-vol.I-41.pdf (2005)

\bibitem{wu05}
Wu B et al. \emph{Crystal Growth \& Design} \textbf{5} 1491 (2005)

\bibitem{wias}
Dreyer W et al. \emph{Milan J. Math.}  \textbf{80} 311   (2012)

\bibitem{huang}
Huang Y "Magentic control in crystal growth from a melt". Ph.D. thesis (Houston: Rice University, 2012)

\bibitem{griesse}
Griesse R, Meir \emph{Math. Comput. Model. Dynam. Syst.} \textbf{15} 163 (2009)

\bibitem{ivan}
Ivanov A et al. \emph{J. Crystal Growth}   \textbf{250} 183 (2003)

\bibitem{meir}
Meir A, Schmidt P \emph{SIAM J. Numer. Anal.} p. 659 (1999)

\bibitem{duffar}
Duffar T  \emph{J. Optoelectron. Advanced Mater.} \textbf{2} 432 (2000)

\bibitem{valek}
V\'alek L, \^Sik J, in \emph{Modern Aspects of Bulk Crystal and Thin Film
  Preparation} (Ed.  N Kolesnikov) (InTech, 2012) p. 45

\bibitem{mullin}
Mullin J \emph{J. Crystal Growth} \textbf{264} 578 (2004)

\bibitem{jin}
Jin C  \emph{J. Semicond.} \textbf{34} 063005 (2013)

\bibitem{karlb}
Carlberg T, King T, Witt A \emph{Solid state sci. technol.} \textbf{120} (1982)

\bibitem{hir}
Hirata H, Hoshikawa K \emph{J. Crystal Growth} \textbf{106} 657 (1990)

\bibitem{abe}
Abe K et al. \emph{Journal of Crystal  Growth} \textbf{181} 41 (1997)

\bibitem{luo02}
Luo B et al. \emph{ Solid-State Electron.}   \textbf{46} 573 (2002)

\bibitem{herro}
Herro Z D et al. \emph{Mater. Res. Soc.  Symp. Proc.} \textbf{892} FF21 (2006)

\bibitem{freitas1}
Slack G A et al. \emph{J. Crystal Growth}   \textbf{246} 287 (2002)

\bibitem{freitas}
Freitas J A  \emph{J. Ceramic Process. Res.} \textbf{6} 209 (2005)

\bibitem{schultz}
Schultz T. "Defect analysis of aluminum nitride". Dr. rer. nat. thesis (Berlin: Technishen Universit\"at,  2010)

\bibitem{kazan1}
Kazan M et al. \emph{Diamond Rel. Mater.} \textbf{15}   1525 (2006)

\bibitem{vaudo}
Vaudo R P et al. \emph{phys. stat. sol. (a)}   \textbf{200} 18 (2003)

\bibitem{p13}
Paskova T et al. \emph{phys. stat. sol.} p. S344  (2009)

\bibitem{78}
Miyazaki M \emph{J. Crystal Growth} \textbf{236} 455 (2002)

\bibitem{80}
Sch\"afer M et al. \emph{Comp. Mater. Sci.}   \textbf{24} 409 (2002)

\bibitem{81}
Tsaur S, Kon S  \emph{J. Crystal Growth} \textbf{249} 470 (2003)

\bibitem{82}
Tsai C  \emph{J. Cryst. Growth} \textbf{113} 499 (1991)

\bibitem{zhm_str}
Zhmakin A I, arXiv:1102.5000 (2011)

\bibitem{83}
Maroudas D, Brown R \emph{J. Crystal Growth} \textbf{108} 399 (1991)

\bibitem{84}
Suezawa M, Sumino K, Yonenaga N  \emph{Phys. Stat. Sol.}  \textbf{A 51} 217 (1979)

\bibitem{85}
Binoiu O et al. \emph{Crystal Res. Technol.} \textbf{36} 707 (2001)

\bibitem{86}
Voronkov V, Falster R J \emph{Electrochem. Soc.} \textbf{149} G167 (2002)

\bibitem{97}
Roache P \emph{Verification and Validation in Computational Science and
  Engineering} (Albuquerque, New Mexico: Hermosa Publishers, 1998)

\bibitem{100}
Oberkampf W, Trucano T \emph{Progr. Aerosp. Sci.} \textbf{38} 209 (2002)

\bibitem{101}
Hatton L  \emph{IEEE Comp. Sci. Eng.} \textbf{4} 27 (1997)

\bibitem{ms1}
Salari P, Knupp K. 
Sandia Report SAND2000-1444 (Sandia Nat. Lab., 2000)

\bibitem{ms2}
Roache P J  \emph{Trans. ASME, J. Fluids Engineering} \textbf{124} 4 (2002)

\bibitem{ms3}
Grier B et al.  \emph{J. Comput. Phys.}  \textbf{278} 193 (2014)

\bibitem{lid}
Ghia U, Ghia K N, Shin C  \emph{J. Comput. Phys.} \textbf{48} 387 (1982)

\bibitem{lid1}
Botella O, Peyret R  \emph{Comput. Fluids} \textbf{27} 421 (1998)

\bibitem{lid4}
Erturk E \emph{Int. J. Numer. Meth. Fluids} \textbf{60} 275 (2009)

\bibitem{102}
De~Vahl~Davis G  \emph{Int. J. Numer. Methods Fluids} \textbf{3} 249 (1983)

\bibitem{103}
Hortmann M, Peri\^c M Scheuerer G \emph{Int. J. Numer. Methods Fluids}
  \textbf{11} 189 (1990)

\bibitem{nc2}
Gjesdal T, Wasberg C, Reif B,
in \emph{CD-ROM Proc. Third Int. Symp.  Adv. Comput.  Heat Transfer} (Begell, 2004)
 CHT-04-172 

\bibitem{nc3}
Wakashima S, Saitoh T \emph{Int. J. Heat Mass Transfer} \textbf{47} 853 (2004)

\bibitem{104}
Blackwell B, Pepper D (eds.), \emph{Benchmark Problems for Heat Transfer
  Codes}   (N.Y.: Am. Soc. Mech. Engin., 1992)

\bibitem{cz1}
Wheeler A \emph{J. Crystal Growth} \textbf{102} 691 (1990)

\bibitem{chen08a}
Chen Q et al.  \emph{Progr. Natur. Sci.} \textbf{18} 1465 (2008)

\bibitem{105}
Amberg G, in \emph{Proc. TSFP-2 Conference, Stockholm, Sweden} (2001)

\bibitem{106}
Zeng Z et al. \emph{J. Crystal Growth} \textbf{229} 601 (2001)

\bibitem{char}
Brown P, in  [20] p. 343

\bibitem{107}
Sunagawa I, in \emph{Crystal Growth Technology} (Eds K Byrappa, T Ohashi) 
  (William Andrew Publishing, 2003), p. 1

\bibitem{108}
Yau S T, Vekilov P \emph{Nature} \textbf{406} 494 (2000)

\bibitem{109}
Vekilov P,  Alexander J I D  \emph{Chem. Rev.}  \textbf{100} 2061 (2000)

\bibitem{110}
McPherson A et al. \emph{J. Structur. Biology}   \textbf{142} 33 (2003)

\bibitem{g5}
Grzegory I et al. \emph{J. Crystal Growth} \textbf{246} 177 (2002)

\bibitem{interf}
Рашкович Л Н,  Шустин О А \emph{УФН} \textbf{151}, 529 (1987)

\bibitem{113}
Stewart M, Weinberg C  \emph{Trans. AIME} \textbf{245} 2108 (1969)

\bibitem{115}
Koster J, Seidel T, Derebai R \emph{J. Fluid Mech.} \textbf{343} 29 (1997)

\bibitem{115a}
Kishida Y, Tanaka M \emph{ Nippon Steel Tech. Rep.} (83) (2001)

\bibitem{116}
Derebail R, Koster J \emph{Int. J. Heat Mass Transfer} \textbf{41} 2537 (1998)

\bibitem{117}
Kakimoto K et al. \emph{J. Crystal Growth} \textbf{99}   665 (1990)

\bibitem{118}
Golyshev V, Gonik M \emph{J. Crystal Growth} \textbf{262} 212 (2004)

\bibitem{119}
Barz R O, Grassal M \emph{J. Crystal Growth} \textbf{249} 345 (2003)

\bibitem{121}
Levenstam M, Anderson M \emph{J. Crystal Growth} \textbf{158} 224 (1996)

\bibitem{122}
Nakamura S, Hibiya T \emph{J. Crystal Growth} \textbf{186} 85 (1998)

\bibitem{123}
M\"uller G et al. \emph{Microel.  Eng.} \textbf{1} 135 (1999)

\bibitem{wadley}
 Wadley H N G, Dharmasea K \emph{J. Crystal Growth} \textbf{130}, 553 (1993)

\bibitem{glazov}
Glazov V, Chizevskaya S, Glagoleva N \emph{Liquid Semiconductors} (N.Y.: Plenum, 1969)

\bibitem{124}
Hintz P, Schwabe D, Wilke H \emph{J. Crystal Growth} \textbf{222} 343 (2001)

\bibitem{pol01}
Polezhaev V et al.  \emph{J. Crystal Growth}   \textbf{230} 40 (2001)

\bibitem{son}
Son S, Nam P O, Yi K W \emph{J. Crystal Growth} \textbf{292} 272 (2006)

\bibitem{125}
Gorbunov L et al. \emph{J. Crystal   Growth} \textbf{257} 7 (2003)

\bibitem{134}
Lin K et al.  \emph{J. Crystal Growth} \textbf{237-239} 1736 (2002)

\bibitem{135}
Eldred M et al. \emph{AIAA-96-4164}  (1996)

\bibitem{136}
Makarov Y, Zhmakin A, \emph{J. Crystal Growth} \textbf{100} 646 (1990)

\bibitem{138}
Zabaras N in \emph{Computational Methods in Optimal Design and Control}
  (1998) p. 391

\bibitem{139}
Jeong J, Kang I \emph{J. Comput. Phys.} \textbf{177} 284 (2002)

\bibitem{140}
Papalambros P, in \emph{Proc. 3rd Int. Symp. Tools Meth. Compet. Eng., Delft} (2000) 14

\bibitem{141}
Giesing J,  Barthelemy J F \emph{AIAA-98-4737}  (1998)

\bibitem{142}
Batill S, Renand J, Gu X \emph{AIAA-2000-4803} (2000)

\bibitem{87}
C{F}{D} {C}odes list.
\newblock http://www.icemcfd.com/cfd/CFD\_codes.htm

\bibitem{88}
Hirsh C \emph{Network Bulletin} \textbf{1} 4 (2001)

\bibitem{126}
Hooks I, in \emph{Proc. 3rd Int. Symp.  NCOSE} (1993)   http://www.incose.org/rwg/writing.html

\bibitem{130}
Steinbrenner J, Wyman N, Chawner J, in \emph{Proc. 9th Int.l Meshing Roundtable} (Sandia Nat.
  Lab., 2000) p. 33

\bibitem{131}
Butlin G, Stops C, in \emph{Proc. 9th Int.l Meshing Roundtable} (Sandia Nat.
  Lab., 1996) p. 7

\bibitem{132a}
Sheng X, Meier I R \emph{IEEE Comput. Graphics Appl.} \textbf{15} 35 (1995)

\bibitem{132}
Weihe K, Willhalm T \emph{Konstanzer Schrift. Math. Inform.} (61)  (1998)

\bibitem{132b}
Barequet G, Sharir M \emph{Comput. Aided Geom. Design} \textbf{12} 207 (1995)

\bibitem{133}
Transcendata {E}urope {L}td. {H}omepage.
\newblock http://www.fegs.co.uk/

\bibitem{habashi}
Habashi W et al. \emph{Int. J. Numer. Meth. Fluids} \textbf{32} 725 (2000)

\bibitem{127}
Zhmakin A \emph{Comm. Num. Meth. Eng.} \textbf{13} 219 (1997)

\bibitem{128}
Prudhomme S et al. \emph{Int. J. Num. Meth.  Eng.} \textbf{56} 1193 (2003)

\bibitem{129}
Sch\"onauer W  \emph{Math. Comput. Simul.} \textbf{54} 269 (2000)

\bibitem{dri}
Drikakis D, Schreck E  \emph{AIAA--93--0062} (1993)

\bibitem{dur}
Durst F et al. \emph{Notes Numer. Fluid Dyn.}   \textbf{38} 79 (1992)

\bibitem{m1:levin}
Levin E, Zhmakin A  \emph{Lect. Notes Comput. Sci.} \textbf{1156} 270 (1996)

\bibitem{m1:paral}
Lukanin D, Kalaev V, Zhmakin A \emph{Lect. Notes Comput. Sci.} \textbf{3019} 469   (2004)

\bibitem{zp}
Zhurakovsky V, Pokholkov Y, Agranovich B L  \emph{Global J. of Engng. Educ.}
  \textbf{5} 7 (2001)

\bibitem{nae}
\emph{Education the engineer of 2020: adapting engineering education to the new
  century} (Washington: The National Academy Press, 2005)

\bibitem{pra}
Prados J, Peterson G, Lattuca L  \emph{J. Eng. Educat.} \textbf{94} 165 (2005)

\bibitem{zyw}
Zywno M, Waalen J \emph{Global J. of Engng. Educ.} \textbf{6} 35 (2002)

\bibitem{it}
\emph{Information Technology (IT)-based educational materials. Workshop Report with
  Recommendations} (Washington: The National Academy Press, 2003)

\bibitem{ed1}
Bogdanov M et al., arXiv:physics/0612184 [physics.ed-ph] (2006)

\bibitem{ed2}
Горбачев Ю Е и др. \emph{Телекоммуникации и  информатизация образования} (5) (2006)

\bibitem{98}
\emph{Glossary of verification and validation terms} 
\newblock http://www.grc.nasa.gov/WWW/ wind/ valid/ tutorial/ glossary.html

\bibitem{h99}
Oden J T  \emph{IACM Express}  (12), 12 (2002)

\bibitem{rs2}
Su W et al. \emph{J. Crystal Growth} \textbf{312} 495 (2010)

\bibitem{rs3}
Ren B et al.  \emph{J. Semicond.} \textbf{26}   1764 (2005)

\bibitem{rs4}
Ren B, Zhao L, Wang H \emph{Rare Metals} \textbf{25}, 7 (2006)

\bibitem{rs5}
Ginkin V et al. \emph{J. Crystal Growth}   \textbf{275}, e121 (2005)

\bibitem{rs6}
Rojo J, Diguez E, Derby J \emph{J. Crystal Growth} \textbf{200} 329 (1999)

\bibitem{derby}
Derby J, in \emph{14th Int. Summer School, Dalian, China)} http://www.
  isscg14.org.cn/attachments/lecture/Derby-ISSCG-14.pdf (2010)

\bibitem{faiez}
Faiez R, Mashhoudi M, Najafi F  \emph{Int. J. Math. comp. Phys. Quantum. Eng.}   \textbf{8} 725 (2014)

\end{thebibliography}
}
\end{document}